\documentclass[11pt]{article}

\usepackage[T1]{fontenc}
\usepackage[utf8]{inputenc}
\usepackage[margin=1in]{geometry}
\usepackage{graphicx}
\usepackage{amsmath,amssymb,amsfonts}
\usepackage{amsthm}
\usepackage[title]{appendix}
\usepackage{booktabs}
\usepackage{subcaption}
\usepackage{placeins}
\usepackage[numbers,sort&compress]{natbib}
\usepackage[hidelinks]{hyperref}

\theoremstyle{plain}

\theoremstyle{remark}

\theoremstyle{definition}

\begin{document}

\title{Advanced Calibration Analysis and Tools: Identifying Influential Observations in Stochastic Interest Rate Model Calibration}

\author{
Philipp Mahler\textsuperscript{1,2,*} \quad Peter Ruckdeschel\textsuperscript{1,3}\\[0.8ex]
\small \textsuperscript{1}Department of Financial Mathematics, Fraunhofer Institute for Industrial Mathematics ITWM\\
\small Kaiserslautern, Germany\\
\small \textsuperscript{2}Department of Mathematics, RPTU Kaiserslautern-Landau, Kaiserslautern, Germany\\
\small \textsuperscript{3}Institute for Mathematics, School of Mathematics and Science, University of Oldenburg\\
\small Oldenburg, Germany\\
\small \textsuperscript{*}Corresponding author: \texttt{philipp.mahler@itwm.fraunhofer.de}
}

\date{}

\maketitle

\begin{abstract}
The accurate calibration of interest rate models is central to market-consistent valuation and Economic Scenario Generators (ESGs). Traditional calibration methods for multi-factor models such as the G2++ model often rely on point estimates, neglecting the influence of specific market data and the quantification of estimation uncertainty. This paper develops a diagnostic framework embedding the calibration problem into non-linear regression theory. It shows that the common industry practice of minimizing the Root Mean Squared Relative Error (RMSRE) is equivalent to a Weighted Least Squares (WLS) problem. This equivalence yields the corresponding formulations for diagnostic tools, including the Weighted Hat Matrix for leverage analysis, Influence Functions for local sensitivity diagnostics, and the Functional Delta Method for local, boundary-respecting confidence intervals. The implementation uses an efficient Jacobian factorization that exploits the analytical tractability of At-The-Money (ATM) caps. The framework is applied to a dataset of Euro ATM caps covering the period 2016--2025. Our empirical analysis reveals a boundary-dominated leverage profile, repeated losses of effective dimensionality due to active parameter constraints, and a diagnostic regime shift in local parameter stability around the post-2022 market transition. The resulting message for actuarial model governance is that low RMSRE is not sufficient for calibration validation. We conclude by discussing the framework's applicability to general least-squares problems while highlighting the computational challenges for instruments lacking closed-form gradients, such as swaptions.
\end{abstract}


\bigskip

\section{Introduction}

\subsection{Motivation and Problem Statement}
The calibration of interest rate models is a fundamental task in financial mathematics applications. It forms the basis for pricing and risk management of many different derivatives. While multi-factor models such as the G2++ are flexible enough to capture complex term structure dynamics, their calibration is a difficult statistical estimation problem.

Calibration routines usually focus on minimizing a goodness-of-fit metric to obtain a single set of optimal parameters. However, the choice of objective function itself requires critical examination. The industry standard, the Root Mean Squared Relative Error (RMSRE), is often used as a default choice without explicitly discussing its statistical interpretation. As discussed by Gneiting \cite{gneiting2011making}, the choice of a loss function links the estimator to a specific functional of the error distribution. In addition, the RMSRE can become numerically unstable when market prices $y^{\text{Mkt}}$ are close to zero.

Furthermore, treating calibration strictly as a numerical optimization problem yields a point estimate without context. In complex models such as the G2++, the non-linear relationship between parameters implies that the objective function may exhibit flat valleys, where different parameter sets produce nearly indistinguishable market prices. Consequently, a satisfactory calibration framework requires more than goodness-of-fit alone; it demands a careful analysis of parameter dependence, numerical stability, and the economic plausibility of the resulting estimates.

Standard calibration practice typically does not address this layer of model risk. In particular, it does not reveal how strongly the result depends on individual market quotes, whether there are issues with parameter identifiability, or how uncertain the fitted parameters remain even when the price fit is numerically good.

This paper links numerical optimization with statistical diagnostics. We argue that a diagnostic calibration framework should answer the following diagnostic questions:
\begin{description}
	\item \textbf{Parameter Identifiability:} Can the model uniquely identify all five parameters, or do optimization constraints reduce the effective degrees of freedom?
	\item \textbf{Geometric Leverage:} Which market instruments drive the local calibration geometry? Is the result determined by only a few boundary maturities?
	\item \textbf{Realized Influence:} To what extent do slight changes in individual input market prices move the calibrated parameter set?
	\item \textbf{Parameter Uncertainty:} How precise are the calibrated values, and how can this uncertainty be quantified in a boundary-respecting way?
\end{description}
Answering these questions is necessary if calibration results are intended to be statistically stable and economically meaningful in addition to being numerically close to market prices. They also provide the organizing logic of the paper: the subsequent sections develop the framework and then answer them empirically for the Euro cap market.

\subsection{Literature Review}\label{subsec:literature_review}

The theory of affine term structure models is well established. Duffie and Kan \cite{duffie1996yield} provide a foundational characterization of affine models, while Filipovi{\'c} \cite{filipovic2009term} develops the associated consistency conditions. For practical implementation, Brigo and Mercurio \cite{brigoMercurio2001} present the standard analytical framework for the G2++ model, including closed-form pricing formulas for interest rate caps.

Despite the emergence of more complex models, the G2++ model remains highly relevant in current regulatory settings such as Solvency~II and PRIIPs, where market-consistent valuation is often required. Applications including the PIA basic model used by the German \textit{Produktinformationsstelle Altersvorsorge} (PIA) and the open-source internal model \textit{openIRM} \cite{wolf2026openirm} continue to rely on a calibrated G2++ model as part of economic scenario generators. Its continued popularity stems from the balance between analytical tractability and the ability to capture complex humped volatility structures that simpler one-factor models cannot reproduce. Precisely because the model remains tractable and operationally relevant, it provides a natural benchmark for the development of calibration diagnostics.

In practice, calibration is often treated primarily as a numerical optimization problem. Mathematically, it can be viewed as an \textit{ill-posed inverse problem} \cite{cont2004financial}, where small data perturbations may lead to parameter instability. Part of the literature addresses this by focusing on the efficiency and stability of optimization algorithms; for instance, Rainer \cite{rainer2009calibration} discusses the general structure of these calibration problems, while Karlsson et al.\ \cite{karlsson2017calibrating} demonstrate the use of Levenberg--Marquardt-type schemes in a related calibration context.

The statistical properties of the calibration estimator appear to be less frequently explored in the practitioner-oriented literature. For example, Christoffersen and Jacobs \cite{christoffersen2004importance} show that the choice of loss function can materially affect parameter estimates, yet many applications continue to rely on point estimates without an explicit probabilistic interpretation. The mathematical distinction between estimation and calibration is fundamental: Desmettre and Korn \cite{desmettre2018moderne} carefully distinguish historical estimation under \(\mathbb{P}\) from implied calibration under \(\mathbb{Q}\). Calibration to option prices identifies a market-consistent risk-neutral specification appropriate for valuation and hedging, whereas historical estimation targets real-world dynamics and long-run scenario behaviour. For actuarial applications, both perspectives are relevant and should be viewed as complementary rather than competing.

Furthermore, while the systematic use of non-linear regression diagnostics \cite{seber1989nonlinear} is standard in other fields, it has rarely been adapted to the calibration of multi-factor interest rate models. The interpretation of estimators as statistical functionals is foundational in robust statistics \cite{huber1981robust,hampel1986robust,reeds1976definition} and is developed in an asymptotic direction by Rieder \cite{rieder1994robust}, yet these tools are only infrequently used to quantify parameter uncertainty, leverage, or local instability in financial calibration.

Conceptually, this adds an interpretability layer to the calibration problem: leverage indicates which market quotes are geometrically important for the local fit, while influence indicates which quotes actually move the calibrated parameters. The framework is therefore closer to analytic, model-specific interpretability than to generic black-box explanation.

The paper therefore interprets the industry-standard RMSRE objective within a Weighted Least Squares framework and derives information-matrix-based uncertainty measures together with leverage and influence diagnostics for the G2++ model.

\subsection{Research Gap and Contribution}
Even though multi-factor models are widely used in applications such as Solvency~II, PRIIPs, and the valuation of German pension products, standard calibration output often reports only the final objective value and point estimate. A central gap is the lack of a transparent statistical framework that diagnoses the local geometry of the calibration problem.

We therefore view calibration as a non-linear statistical problem. Our contributions are as follows:

\begin{description}
	\item[Statistical--Geometric Interpretation via WLS.] 
	Interpreting the calibration as a Weighted Least Squares (WLS) problem links the RMSRE objective to regression diagnostics and gives a geometric interpretation of the calibration problem. The RMSRE objective determines how distances between market prices and model prices are measured. This, in turn, defines the relevant local geometry of the G2++ model surface and its tangent approximation. Conversely, this geometry provides statistical information: the tangent space describes local identifiability, the Weighted Hat Matrix measures geometric leverage, the Fisher-information-type curvature quantifies local parameter uncertainty, and Influence Functions combine geometric sensitivity with the realized residuals. This establishes a bridge between statistical diagnostics and the local information-geometric interpretation of the G2++ pricing map. Since the pricing map is non-linear and the calibration is constrained, these diagnostics cannot be used off the shelf but have to be formulated locally, conditionally on the realized weights, and in terms of the weighted Jacobian of the pricing map.
	
	\item[Efficient Implementation via Jacobian Factorization.] 
	Diagnostic tools require the Jacobian of the pricing function. Computing numerical derivatives for the entire portfolio is costly and can be unstable. We derive an analytical factorization of the Jacobian for G2++ caps, separating the market-data-dependent factor from the reusable variance sensitivities and thereby making the implementation computationally efficient.
	
	\item[Uncertainty Quantification and Stability Analysis.] 
	Standard normal approximations are often problematic because interest rate parameters are subject to natural boundaries (e.g., $|\rho_{xy}| \le 1$). We therefore combine the \textit{Functional Delta Method} with Variance Stabilizing Transformations (VST) to obtain boundary-respecting confidence intervals. Furthermore, we introduce \textit{Effective Degrees of Freedom} diagnostics to detect local dimensionality collapses caused by rank deficiency or active constraints.
\end{description}

Our diagnostic framework is not a full economic or actuarial validation of the calibrated model. The quantities studied here are local diagnostics of the calibration geometry of the market fit at $t=0$. They reveal whether the parameter vector is stably identified, which instruments dominate the fit, and how much parameter uncertainty remains even when the price fit is numerically good; but they do not by themselves certify long-run simulation behaviour.

For actuarial model governance, this diagnostic layer is useful because ESG calibrations enter market-consistent valuation, solvency applications, and pension or insurance projections. In practical applications, it should therefore be complemented by a second layer of ex-post validation, including the behaviour of simulated short-rate paths, the long-run slope of the yield curve, or the share of inverted term structures after calibrating the relevant risk premia. The framework thus provides a transparent local check before calibrated parameters are used for valuation or scenario generation, but not the entire validation problem.

The empirical part is organized around this distinction. We use three economically recognizable dates as \emph{temporal landmarks}. In this paper, this means that the same dates are used as common reference points across the diagnostic plots. They are not interpreted as a separate event study. Their role is to make the diagnostic layers comparable: days with similarly low RMSRE can differ substantially in leverage, effective dimensionality, and parameter uncertainty.

\subsection{Outline of the Paper}
The paper is structured as follows. Section~2 establishes the financial mathematics setup of the G2++ model. Section~3 develops the statistical framework, proves the WLS equivalence, and derives the analytical diagnostics. Section~4 describes the data processing and the efficient computational implementation. Section~5 answers the four diagnostic questions empirically by analyzing fit quality, leverage and influence, effective dimensionality, regime-dependent calibration geometry, and parameter uncertainty over time. Section~6 concludes with implications for actuarial practice. Detailed mathematical theory and derivations are collected in the Appendix.

\section{Setup and Mathematical Formulation}
\label{sec:setup}

In this section, we define the mathematical components of the calibration problem. We follow the notation and framework established in Brigo and Mercurio \cite{brigoMercurio2001} for the G2++ model. We begin by defining the stochastic properties of the model and deriving the pricing formulas for the calibration instruments. This setup forms the basis for the statistical framework developed in Section~\ref{sec:theoretical_framework}.

\subsection{The G2++ Interest Rate Model}
The G2++ model characterizes the instantaneous short rate $r(t)$ as the sum of two correlated Ornstein-Uhlenbeck processes and a deterministic function. Under the risk-neutral measure $\mathbb{Q}$, the dynamics are given by:
\begin{equation}
	r(t) = x(t) + y(t) + \psi(t),
\end{equation}
where the stochastic processes $x(t)$ and $y(t)$ satisfy:
\begin{align}
	dx(t) &= -a_x x(t)dt + \sigma_x dW_1(t), \quad x(0) = 0, \\
	dy(t) &= -a_y y(t)dt + \sigma_y dW_2(t), \quad y(0) = 0.
\end{align}
The Brownian motions $W_1(t)$ and $W_2(t)$ are two standard $\mathbb{Q}$-Brownian motions with correlation $d\langle W_1, W_2 \rangle_t = \rho_{xy} dt$. The deterministic function $\psi(t)$ ensures that the model perfectly fits the observed initial bond prices at time $t=0$:
\begin{equation}\label{eq:psi}
	\psi(t) = f^M(0,t) + \frac{\sigma_x^2}{2a_x^2} \left( 1 - e^{-a_x t} \right)^2 + \frac{\sigma_y^2}{2a_y^2} \left( 1 - e^{-a_y t} \right)^2 + \rho_{xy} \frac{\sigma_x \sigma_y}{a_x a_y} \left( 1 - e^{-a_x t} \right) \left( 1 - e^{-a_y t} \right).
\end{equation}

The explicit solutions for $x(t)$ and $y(t)$ are given by:
\begin{align}
	x(t) &= \sigma_x \int_0^t e^{-a_x (t-u)} \, dW_1(u), \\ 
	y(t) &= \sigma_y \int_0^t e^{-a_y (t-u)} \, dW_2(u).
\end{align}
Consequently, both processes are normally distributed with:
\begin{equation}
	x(t) \sim \mathcal{N}\left(0, \sigma_x^2 \frac{1-e^{-2a_x t}}{2a_x} \right),\quad y(t) \sim \mathcal{N}\left(0, \sigma_y^2 \frac{1-e^{-2a_y t}}{2a_y} \right).
\end{equation}
Hence, under the risk-neutral measure $\mathbb{Q}$, the expected value of the short rate is given by $\mathbb{E}_{\mathbb{Q}}\left[ r(t)\right]=\psi(t)$, whereas the variance is given by:
\begin{align}
	\text{Var}\left[ r(t)\right] &= \frac{\sigma_x^2}{2a_x} \left(1-e^{-2a_x t}\right)+ \frac{\sigma_y^2}{2a_y}\left(1-e^{-2a_y t}\right) + 2 \rho_{xy} \frac{\sigma_x\sigma_y}{a_x+a_y} \left(1-e^{-\left(a_x+a_y\right)t}\right).
\end{align}
A key feature of the model is that zero-coupon bond prices under the risk-neutral measure can be expressed in closed form (see \cite{brigoMercurio2001}):
\begin{equation}
	\label{eq:ZeroBondPrice}
	P(t,T) = \exp\left\{ - \int_t^T {\psi(u) du}  - \frac{{1 - {e^{ - a_x\left( {T - t} \right)}}}}{a_x} x(t)  - \frac{{1 - {e^{ - a_y\left( {T - t} \right)}}}}{a_y} y(t) + \frac{1}{2}V(t,T) \right\}
\end{equation}
with the auxiliary function $V(t,T)$ defined as: 
\begin{align}
	V(t,T) &= \frac{\sigma_x ^2}{a_x^2}\left[ {T - t + \frac{2}{a_x}{e^{ - a_x\left( {T - t} \right)}} - \frac{1}{2a_x} {e^{ - 2a_x\left( {T - t} \right)}} - \frac{3}{2a_x}} \right]  \nonumber \\
	&+ \frac{\sigma_y ^2}{a_y^2}\left[ {T - t + \frac{2}{a_y}{e^{ - a_y\left( {T - t} \right)}} - \frac{1}{2a_y}{e^{ - 2a_y\left( {T - t} \right)}} - \frac{3}{2a_y}}\right] \nonumber \\
	&+ 2\rho_{xy}\frac{\sigma_x \sigma_y}{a_x a_y}\left[ {T - t + \frac{e^{ - a_x\left( T - t \right)} - 1}{a_x} + \frac{e^{ - a_y\left( T - t \right)} - 1}{a_y} - \frac{e^{ - \left( a_x + a_y \right)\left( T - t \right)} - 1}{a_x + a_y}} \right].
\end{align}
The parameter vector to be calibrated is $\Pi = (a_x, a_y, \sigma_x, \sigma_y, \rho_{xy})^\top$.

\subsection{Calibration Instruments: Interest Rate Caps}
\label{subsec:cap_pricing}
We use At-The-Money (ATM) interest rate caps for calibration. They are popular derivatives for hedging against interest rate risk and are hence frequently traded.

A cap is a series of consecutive European call options (known as \textit{caplets}) on a floating reference rate, such as EURIBOR. It protects the buyer from a rise in the reference rate above an agreed-upon exercise price (strike) $K$. The payoff of a single caplet at time $t_i$ for the interest period $[t_{i-1}, t_i]$, with period length $\delta_i = t_i - t_{i-1}$ and notional $N$, is:
\begin{equation}
	\label{eq:caplet_payoff}
	\text{Payoff}_{\text{Caplet}}(t_i) = N \cdot \delta_i \cdot \max(L(t_{i-1}, t_i) - K, 0).
\end{equation}
Here, $L(t_{i-1}, t_i)$ denotes the floating reference rate determined at the fixing date $t_{i-1}$ for payment at $t_i$.

Notably for the G2++ framework, a caplet on the rate $L(t_{i-1}, t_i)$ is mathematically equivalent to a \textit{put option} on a zero-coupon bond $P(t_{i-1}, t_i)$ with bond strike $K^* = \frac{1}{1+K\delta_i}$. This equivalence allows us to use the closed-form bond price formula inherent to the G2++ model; see Eq.~\eqref{eq:ZeroBondPrice}.

Consequently, the price of a cap at $t=0$ for a set of payment dates $\mathcal{T} = \{t_m, \dots, t_n\}$ is given by:
\begin{equation}
	\label{eq:analytical_cap_price}
	\text{Cap}(0,\mathcal{T}, N, K, \Pi) = \sum_{i=m+1}^n N \left[ P(0, t_{i-1}) \Phi(h_{+,i}) - (1 + K \delta_i) P(0, t_i) \Phi(h_{-,i}) \right],
\end{equation}
where $\Phi(\cdot)$ is the cumulative standard normal distribution function and the terms $h_{\pm,i}$ are defined as:
\begin{equation}
	\label{eq:h_pm}
	h_{\pm,i} = \frac{1}{\Sigma(0, t_{i-1}, t_i)} \left[ \log \left( \frac{P(0, t_{i-1})}{(1 + K \delta_i) P(0, t_i)} \right) \pm \frac{1}{2} \Sigma(0, t_{i-1}, t_i)^2 \right].
\end{equation}
The integrated variance function $\Sigma^2$ is derived directly from the bond price variance structure:
\begin{align}
	\Sigma(0, T_1, T_2)^2 &= \sigma_x^2 C(a_x, T_1, T_2)^2 C(2 a_x, 0, T_1) + \sigma_y^2 C(a_y, T_1, T_2)^2 C(2 a_y, 0, T_1) \nonumber \\
	&\quad + 2 \rho_{xy} \sigma_x \sigma_y C(a_x, T_1, T_2) C(a_y, T_1, T_2) C(a_x + a_y, 0, T_1),
\end{align}
using the auxiliary function $C(\alpha, t, T) = \frac{1-e^{-\alpha(T-t)}}{\alpha}$.

A cap is referred to as \textit{at-the-money} (ATM) when its strike $K$ corresponds to the fixed rate of a newly issued interest rate swap with an identical maturity and interest period. For a cap starting today ($t=0$), the ATM strike is therefore the current market swap rate.

\subsection{The Optimization Problem and Numerical Solution}
\label{subsec:optimization}
To use the model in practice, it must be linked to observable financial market data. This link is established through the determination of the model parameters $\Pi$, a process generally referred to as calibration. The objective is to find the estimator $\hat{\Pi}$ that minimizes the discrepancy between model prices $\text{Cap}(0,\mathcal{T}_k, N_k, K_k, \Pi)$ and market prices $y^{\text{Mkt}}_k$. To avoid bias towards high-priced instruments (long maturities), which would dominate a simple sum of squared errors, we minimize the Root Mean Squared Relative Error (RMSRE). For a set of $m$ different instruments, the optimization problem is of the following form:

\begin{equation}
	\label{eq:rmsre_objective}
	\begin{aligned}
		& \hat{\Pi} = \arg\min_{\Pi}  \sqrt{\frac{1}{m} \sum_{k=1}^{m} \left( \frac{y^{\text{Mkt}}_k - \text{Cap}(0,\mathcal{T}_k, N_k, K_k, \Pi)}{y^{\text{Mkt}}_k} \right)^2} \\
		& \text{subject to:} \\
		& \qquad \begin{aligned}
			a_x, a_y, \sigma_x, \sigma_y & > 0, \\
			-1 \leq \rho_{xy} & \leq 1.
		\end{aligned}
	\end{aligned}
\end{equation}

This optimization is typically performed using numerical algorithms that can broadly be categorized as \textit{gradient-free} and \textit{gradient-based} approaches. Gradient-free methods -- including stochastic algorithms such as Adaptive Simulated Annealing (ASA) or local, model-based sol\-vers such as BOBYQA and Nelder--Mead -- are often preferred for the G2++ model to ensure robustness against local minima in complex objective landscapes. While these methods determine the optimal parameters without requiring derivatives, the analytical Jacobian remains essential for the subsequent diagnostic analysis. In contrast, gradient-based methods such as the Levenberg--Marquardt algorithm use derivative information directly during the search process to accelerate convergence. In this study, we use gradient-free solvers for the point estimation, while the derivative-based framework developed in Section~\ref{sec:theoretical_framework} is employed to analyze the stability and uncertainty of the resulting calibration.

\subsection{The Challenge of Non-Linearity}
While the optimization problem above provides a point estimate $\hat{\Pi}$, the relationship between model parameters and observable market prices is inherently non-linear, and this non-linearity poses substantial conceptual and numerical challenges.

To illustrate this point, it is useful to contrast the calibration problem with standard linear regression. In a linear setting, model outputs depend linearly on the parameters, and the resulting least-squares problem admits a unique global minimum with well-understood statistical properties. Small changes in the data lead to small and predictable changes in the estimated parameters.

In contrast, the pricing formula derived in Eq.~\eqref{eq:analytical_cap_price} depends non-linearly on the parameter vector $\Pi$. As a consequence, the objective function measuring the deviation between model prices and observed market prices exhibits a much more complex structure. This complexity manifests in several ways:

\begin{description}
	\item[\textbf{Non-linear Parameter--Price Mapping}]
	The G2++ model cannot reproduce every arbitrary combination of cap prices observed in the market. The set of all possible cap price vectors that the model \textit{can} generate forms a curved surface embedded within the space of market prices of caps. Calibration is geometrically equivalent to finding the point on this surface that is closest to the observed market data point. Because the surface is curved, a simple linear projection is only accurate locally.
	
	\item[\textbf{Multiple Local Optima}]  
	Due to the non-linear structure of the pricing functions, the objective function may possess several local minima. Standard gradient-based optimization algorithms can converge to different parameter sets with almost equal objective values.
	
	\item[\textbf{Parameter Interdependence and Flat Directions}]  
	In many cases, different parameters influence market prices in a similar way. For example, an increase in the mean reversion speed of the first factor ($a_x$) can often be compensated by a corresponding adjustment in the second factor ($a_y$) or the volatilities to produce nearly identical derivative prices. This often results in the objective function forming a ``flat valley'' rather than a distinct minimum, leading to numerical instability.
\end{description}

These observations motivate the statistical diagnostic framework developed in the next section.

\section{Calibration as Weighted Least Squares}
\label{sec:theoretical_framework}

A low value of the objective function confirms that the model is capable of reproducing the observed market prices. However, this numerical result provides no insight into the stability of the estimated parameters, their uncertainty, or the influence of individual market data points. To analyze these properties, we use diagnostic tools from statistical regression theory by formalizing the calibration as a non-linear regression problem with fixed design.

We define the regression function, denoted by $g_k(\Pi)$, as the analytical pricing formula derived in Section~\ref{subsec:cap_pricing}:
\begin{equation}
	g_k(\Pi) \equiv \text{Cap}(0,\mathcal{T}_k, N_k, K_k, \Pi).
\end{equation}
We assume that the observed market prices $y^{\text{Mkt}}_k$ are generated by this model subject to a multiplicative error structure, $y^{\text{Mkt}}_k = g_k(\Pi) \cdot (1 + \varepsilon_k)$. This structure implies that pricing errors are proportional to the magnitude of the instrument's price, which provides the statistical justification for minimizing relative errors rather than absolute errors.

In this section, we first prove that the standard RMSRE objective is equivalent to a Weighted Least Squares (WLS) estimator. Building on this equivalence, we use the geometric interpretation of linear models and the Functional Delta Method to derive established diagnostics for parameter uncertainty and sensitivity.

\subsection{Equivalence of RMSRE and Weighted Least Squares}
\label{subsec:wls_equivalence}

To establish a formal link between the calibration objective and regression theory, we show that minimizing the RMSRE defined in Eq.~\eqref{eq:rmsre_objective} is equivalent to a Weighted Least Squares (WLS) estimator. The parameter constraints determine the feasible set, whereas the weighting structure of the objective determines how pricing errors enter the minimization.

Since the square root is a strictly increasing function for non-negative arguments, the location of the minimum is invariant under a square transformation. Thus, minimizing the RMSRE is equivalent to minimizing the Mean Squared Relative Error (MSRE). Furthermore, the constant scaling factor $\frac{1}{m}$ is independent of the parameter vector $\Pi$ and does not affect the argmin. Finally, by algebraically separating the market price from the numerator, we can isolate the pricing error $(y_k^{\text{Mkt}} - g_k(\Pi))$ and interpret the inverse of the squared market price as a weighting factor. This sequence of transformations is summarized as follows:
\begin{equation}
	\label{eq:wls_derivation}
	\begin{aligned}
		\hat{\Pi} &= \arg \min_{\Pi} \left( \sqrt{\frac{1}{m} \sum_{k=1}^{m} \left( \frac{y_k^{\text{Mkt}} - g_k(\Pi)}{y_k^{\text{Mkt}}} \right)^2} \right)^2 \\
		&= \arg \min_{\Pi} \sum_{k=1}^{m} \frac{1}{(y_k^{\text{Mkt}})^2} \left( y_k^{\text{Mkt}} - g_k(\Pi) \right)^2.
	\end{aligned}
\end{equation}

This final expression is the canonical form of a \textit{Weighted Sum of Squared Errors (WSSE)}. Consequently, calibration is equivalent to solving:
\begin{equation}
	\label{eq:wls_form}
	\hat{\Pi} = \arg \min_{\Pi} \sum_{k=1}^{m} w_k \left( y_k^{\text{Mkt}} - g_k(\Pi) \right)^2, \qquad \text{with weights } w_k = \frac{1}{(y_k^{\text{Mkt}})^2}.
\end{equation}

In the subsequent diagnostic analysis, the empirical weights $w_k$ are treated as fixed after observation. Hence, all geometric and asymptotic arguments are understood conditionally on the realized weight matrix $\mathbf{W}$.
From a statistical perspective, the choice of the relative error objective elicits a specific functional of the underlying error distribution. As discussed by Gneiting \cite{gneiting2011making}, while standard MSE targets the conditional mean $\mathbb{E}[Y]$, minimizing relative errors targets the ratio of moments $\mathbb{E}[Y^{-1}] / \mathbb{E}[Y^{-2}]$. Relative-error objectives remain potentially problematic when observed prices are
close to zero, since the induced WLS weights
$w_k=(y_k^{\text{Mkt}})^{-2}$ may then become unstable.
In our empirical cap panel, however, this issue is not active: the smallest market
price entering the calibration objective is $2.6\times 10^{-3}$, and no
calibration price falls below $10^{-4}$. This target is inherently robust to the price scale, thereby assigning comparable importance to relative deviations across maturities. Furthermore, while the weights $w_k$ are data-dependent, this approach is justified under the assumption of a multiplicative error structure, where the market price serves as a proxy for the local variance.

Beyond the theoretical insights, this least-squares structure also provides a computational advantage for numerical optimization. Instead of relying exclusively on general-purpose Newton procedures, one can use the Gauss--Newton algorithm or related variants. These methods exploit the specific form of Eq.~\eqref{eq:wls_form} to avoid the explicit calculation of full second derivatives while still using the local curvature information contained in $\mathbf{J}^\top \mathbf{W} \mathbf{J}$.

By establishing this WLS equivalence, we can apply the asymptotic theory of non-linear regression to derive the covariance matrix and influence functions used in the subsequent sections.

\subsection{The Geometry of Calibration: Tangent Space Projection}
\label{subsec:geometry}
Having established the WLS nature of the problem, we can use the geometric interpretation of linear models to define our diagnostics. Let $\mathbf{W} = \operatorname{diag}(w_1, \dots, w_m)$ denote the diagonal weight matrix from Eq.~\eqref{eq:wls_form}.

In a standard linear regression $\mathbf{y} = \mathbf{X}\beta + \varepsilon$, the design matrix $\mathbf{X}$ spans the linear subspace onto which the observations $\mathbf{y}$ are projected. The operator performing this projection is the \textit{hat matrix} $\mathbf{H} = \mathbf{X}(\mathbf{X}^\top \mathbf{X})^{-1}\mathbf{X}^\top$. This matrix depends exclusively on the design matrix $\mathbf{X}$, meaning the geometry of the projection is fixed and independent of both the observations $\mathbf{y}$ and the parameter estimates $\beta$. From a linear algebra perspective, the hat matrix is constructed using the pseudo-inverse of $\mathbf{X}$, commonly denoted as $\mathbf{X}^+ = (\mathbf{X}^\top \mathbf{X})^{-1}\mathbf{X}^\top$, which identifies the unique solution that minimizes the Euclidean distance to the subspace.

We define the vector-valued pricing function $\mathbf{g}(\Pi) = [g_1(\Pi), \dots, g_m(\Pi)]^\top$, which maps the 5-dimensional parameter space to a curved surface embedded within the $m$-dimensional observation space $\mathbb{R}^m$. To analyze the stability of the calibration, we linearize the model locally around the optimal parameter set $\hat{\Pi}$ using a first-order Taylor expansion:
\begin{equation}
	\label{eq:linearized_model}
	\mathbf{g}(\Pi) \approx \mathbf{g}(\hat{\Pi}) + \mathbf{J}(\hat{\Pi}) \cdot (\Pi - \hat{\Pi}),
\end{equation}
where $\mathbf{J}(\hat{\Pi}) \in \mathbb{R}^{m \times 5}$ is the Jacobian matrix containing the first partial derivatives of the model prices. The diagnostics below are based on this local first-order approximation. Its adequacy is assessed empirically in Appendix~\ref{subsec:gn_adequacy} by comparing the exact objective Hessian with its Gauss--Newton approximation.

\begin{figure}[htbp]
	\centering
	\includegraphics[width=0.5\textwidth]{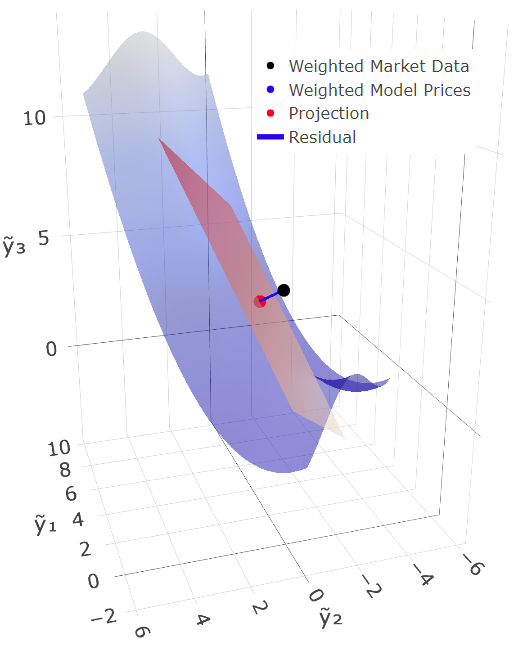}
	\caption{\textbf{Geometric Interpretation.} 
	To visualize the orthogonal geometry of the Weighted Least Squares (WLS) problem, the ambient space represents the \textit{weighted} prices of three instruments, denoted as $\tilde{y}_i = \sqrt{w_i} y_i$. 
	The blue surface represents the weighted non-linear model surface $\tilde{\mathbf{g}}(\Pi) = \mathbf{W}^{1/2}\mathbf{g}(\Pi)$. 
	Calibration geometrically seeks the point on this surface closest to the Weighted Market Data ($\tilde{\mathbf{y}}$, black dot) using standard Euclidean distance (thick blue line). 
	Our diagnostic framework relies on a \textit{local linearization}: the Weighted Hat Matrix $\mathbf{H}_w$ acts as an orthogonal projector, mapping the market data onto the tangent plane (red surface).}
\label{fig:geometry}
\end{figure}

Geometrically, the columns of the Jacobian matrix $\mathbf{J}$ span the \textit{tangent space} to the model surface -- the set of all possible model price vectors $\mathbf{g}(\Pi)$ -- at the solution point. In this local approximation, the Jacobian $\mathbf{J}$ plays exactly the same role as the design matrix $\mathbf{X}$ in linear regression. As illustrated in Figure~\ref{fig:geometry}, the calibration process can be viewed as an orthogonal projection onto this tangent space. To achieve an orthogonal geometry, we consider the coordinate system where all observations are weighted by $\mathbf{W}^{1/2}$. In this transformed space, the Weighted Hat Matrix $\mathbf{H}_w$ acts as the orthogonal projector, and the distance between the market data and the model surface corresponds to the standard Euclidean norm of the relative pricing errors.

We analyze the Jacobian rather than the gradient of the objective function because the Jacobian describes the local parameter-to-price sensitivity. While the first-order optimality conditions identify the location of the estimate $\hat{\Pi}$, they do not describe the local geometry or stability of the solution. To assess parameter uncertainty and leverage, we require information about the curvature of the objective function. In our WLS framework, this curvature is approximated using the product $\mathbf{J}^\top \mathbf{W}\mathbf{J}$, which represents the local sensitivity of the model prices to parameter changes.

\subsection{Analytical Derivation of the Jacobian Matrix}
\label{subsec:jacobian_derivation}

The G2++ model provides a closed-form analytical cap pricing formula; see Eq.~\eqref{eq:analytical_cap_price}. This allows us to derive exact analytical expressions for the Jacobian matrix $\mathbf{J}$ and the full Hessian matrix $\mathbf{H}$. In contrast, the pricing of swaptions in the G2++ model requires a numerical integration. Computing the derivatives of such an integral -- either through finite differences or by differentiating under the integral -- inevitably introduces numerical approximation errors.

By using exact analytical derivatives of caps, the resulting diagnostics reflect the local model geometry without additional finite-difference error. Furthermore, while standard regression diagnostics often rely on the Gauss--Newton approximation $\mathbf{J}^\top \mathbf{W} \mathbf{J}$, deriving the exact Hessian allows for potential use in second-order optimization algorithms (like Newton--Raphson) and supports the local linearity assumption. An empirical comparison between the exact objective Hessian and its Gauss--Newton approximation is reported in Appendix~\ref{subsec:gn_adequacy}, where the residual-weighted second-order correction is found to be negligible throughout our sample.

\paragraph{Chain Rule Along a Path}
The derivative of a single caplet price $\text{Caplet}_i$ (see \eqref{eq:analytical_cap_price}) with respect to a parameter $p$ results from the chain of dependencies:
\[
\text{Caplet}_i \;\;\to\;\; \Phi \;\;\to\;\; h_{\pm} \;\;\to\;\; \Sigma \;\;\to\;\; \Sigma^2 \;\;\to\;\; p.
\]
Formally, this structure translates into the following chain rule expansion:
\begin{equation}
	\label{eq:chain_rule_expansion}
	\frac{\partial \text{Caplet}_i}{\partial p} = \frac{\partial \text{Caplet}_i}{\partial \Phi} \cdot \frac{\partial \Phi}{\partial h} \cdot \frac{\partial h}{\partial \Sigma_i} \cdot \frac{\partial \Sigma_i}{\partial \Sigma_i^2} \cdot \frac{\partial \Sigma_i^2}{\partial p}.
\end{equation}
The explicit derivations for each term in this chain are provided in Appendix~\ref{app:first_order_derivatives}. This formulation highlights that the influence of the parameter $p$ on the caplet price is transmitted exclusively through the integrated variance $\Sigma_i^2$.

\paragraph{Step 1: The Price-to-Variance Sensitivity (\texorpdfstring{$\mathcal{M}_i$}{Mi})}
We first differentiate the caplet price with respect to the variance $\Sigma_i^2$. A direct chain-rule expansion yields a longer expression involving the auxiliary quantity
\[
L_i = \log\left(\frac{P(0,t_{i-1})}{(1+K\delta_i)P(0,t_i)}\right),
\]
but this expression simplifies exactly to the weighting factor
\begin{align}
	\label{eq:factor_M}
	\mathcal{M}_i(\Pi) := \frac{\partial \text{Caplet}_i}{\partial \Sigma_i^2}
	= \frac{N\,P(0,t_{i-1})\varphi(h_{+,i})}{2\Sigma_i}.
\end{align}
The derivation and the exact simplification of the longer chain-rule expression are provided in Appendix~\ref{app:first_order_derivatives}. This factor collects all market-data specific information (strike, maturity, current yield curve) but is independent of the specific parameter being differentiated.

\paragraph{Step 2: The Variance-to-Parameter Sensitivity}
The second component is the derivative of $\Sigma_i$ with respect to the parameters, expressed through the integrated variance function $\Sigma_i^2$:
\[
\frac{\partial \Sigma_i}{\partial p} = \frac{1}{2\Sigma_i} \frac{\partial \Sigma_i^2}{\partial p}.
\]
The explicit analytical expressions for $\frac{\partial \Sigma_i^2}{\partial p}$ for all parameters $p \in \{a_x, a_y, \sigma_x, \sigma_y, \rho_{xy}\}$ are detailed in Appendix \ref{app:final_expressions}.

\paragraph{Step 3: The Factorized Jacobian}
Combining these steps, the entry in the Jacobian matrix for the $k$-th cap (consisting of a set of caplets) with respect to parameter $p$ is:
\begin{equation}
	\label{eq:jacobian_factorization}
	\frac{\partial \text{Cap}_k}{\partial p} = \sum_{i \in \text{Cap}_k} \mathcal{M}_i(\Pi)\cdot\frac{\partial \Sigma_i^2}{\partial p}.
\end{equation}
This factorization allows for a highly efficient implementation: $\mathcal{M}_i$ is computed once per caplet, while the variance sensitivities from Appendix~\ref{app:final_expressions} are reused.

\paragraph{Remark: Generalization to Interest Rate Floors}
While our derivations focus on interest rate caps, the underlying geometric framework extends directly to interest rate floors. By fundamental no-arbitrage pricing, applying the put--call parity to the constituent caplets and floorlets (for a detailed derivation, see Appendix~\ref{app:cap_floor_parity}, and e.g., \cite[Def.~1.6.1]{brigoMercurio2001}) shows that the difference between a cap and a floor with an identical strike $K$ and payment schedule $\mathcal{T} = \{t_m, \dots, t_n\}$ collapses to a payer swap with payment structure $\mathcal{T}$:
\begin{equation}
	\text{Cap}(0, \mathcal{T}, N, K, \Pi) - \text{Floor}(0, \mathcal{T}, N, K, \Pi) = \text{PayerSwap}(0, \mathcal{T}, N, K).
\end{equation}
The present value of this payer swap can be expressed as the sum of its individual cash flows. Exploiting standard no-arbitrage relations between forward rates and zero-coupon bonds, the floating leg forms a telescoping sum, yielding:
\begin{align}
	\text{PayerSwap}(0, \mathcal{T}, N, K) &= \sum_{i=m+1}^n N \Big[ P(0, t_{i-1}) - (1 + K\delta_i)P(0, t_i) \Big] \nonumber \\
	&= N \Big[ P(0, t_m) - P(0, t_n) \Big] - N \cdot K \sum_{i=m+1}^n \delta_i P(0, t_i). \label{eq:swap_telescope}
\end{align}
This swap valuation relies purely on the initial discount curve $P(0,t)$ and is model-independent. In the specific context of the G2++ framework, the initial market term structure is perfectly fitted via the deterministic function $\psi(t)$. Under the standard assumption of a fixed, exogenously provided initial discount curve, Eq.~\eqref{eq:swap_telescope} is completely independent of the model parameters $\Pi = (a_x, a_y, \sigma_x, \sigma_y, \rho_{xy})^\top$. Differentiating both sides of the parity equation with respect to any parameter $p \in \Pi$ thus yields:
\begin{equation}
	\frac{\partial \text{Cap}}{\partial p} - \frac{\partial \text{Floor}}{\partial p} = 0 \quad \implies \quad \mathbf{J}_{\text{Cap}} = \mathbf{J}_{\text{Floor}}.
\end{equation}
For an analytical proof confirming this exact identity at the level of individual caplet and floorlet sensitivities, we refer to Appendix~\ref{app:floor_jacobian}.

Consequently, the price-to-variance sensitivity factor derived in Eq.~\eqref{eq:factor_M} is identical for both instruments ($\mathcal{M}_i^{\text{Cap}} = \mathcal{M}_i^{\text{Floor}}$). 

Furthermore, an instrument is defined as At-The-Money (ATM) precisely when the strike $K$ is chosen to match the forward swap rate, such that the initial present value of the swap in Eq.~\eqref{eq:swap_telescope} is zero. Under arbitrage-free market conditions, this implies identical market prices for ATM caps and floors ($y^{\text{Mkt}}_{\text{Cap}} = y^{\text{Mkt}}_{\text{Floor}}$). Assuming an identical observation and maturity structure, the regression weights in our WLS formulation are exactly the same ($\mathbf{W}_{\text{Cap}} = \mathbf{W}_{\text{Floor}}$). Because both the model geometry ($\mathbf{J}$) and the weighting ($\mathbf{W}$) coincide, the symmetric Weighted Hat Matrix $\mathbf{H}_w = \mathbf{W}^{1/2}\mathbf{J}(\mathbf{J}^\top\mathbf{W}\mathbf{J})^{-1}\mathbf{J}^\top\mathbf{W}^{1/2}$ is invariant. Thus, the diagnostic tools derived in Section~\ref{subsec:diagnostic_tools} for ATM caps apply one-to-one to ATM floors.

\subsection{Derivation of the Hessian Matrix}
\label{subsec:hessian_derivation}

For second-order optimization algorithms and for comparing the exact local curvature with its Gauss--Newton approximation, second-order derivatives are required. At the caplet level, the second derivative
\[
\frac{\partial^2 \text{Caplet}_i}{\partial p_u \partial p_v}
\]
follows by applying the product rule to the factorized first-order representation:
\begin{equation}
	\label{eq:hessian_product_rule}
	\frac{\partial^2 \text{Caplet}_i}{\partial p_u \partial p_v}
	=
	\underbrace{\frac{\partial \mathcal{M}_i(\Pi)}{\partial p_u}\frac{\partial \Sigma_i^2}{\partial p_v}}_{\text{Term 1}}
	+
	\underbrace{\mathcal{M}_i(\Pi)\frac{\partial^2 \Sigma_i^2}{\partial p_u \partial p_v}}_{\text{Term 2}}.
\end{equation}
The corresponding second derivatives for a cap are then obtained by summing over its constituent caplets.

\paragraph{Term 1: Derivative of the Weighting Factor}
The term $\frac{\partial \mathcal{M}_i(\Pi)}{\partial p_u}$ is again obtained via the chain rule, since $\mathcal{M}_i$ depends on $p_u$ exclusively through $\Sigma_i$. We define
\[
\mathcal{M}'_i(\Pi) := \frac{1}{2\Sigma_i}\frac{d\mathcal{M}_i}{d\Sigma_i}.
\]
Differentiating Eq.~\eqref{eq:factor_M} yields the compact representation
\[
\mathcal{M}'_i(\Pi)
=
\frac{N\,P(0,t_{i-1})\varphi(h_{+,i})}{4\Sigma_i^3}
\bigl(h_{+,i}h_{-,i}-1\bigr),
\]
and therefore
\[
\frac{\partial \mathcal{M}_i(\Pi)}{\partial p_u}
=
\mathcal{M}'_i(\Pi)\frac{\partial \Sigma_i^2}{\partial p_u}.
\]
Hence the caplet-level second derivative can be written in the symmetric factorized form
\begin{equation}
	\label{eq:hessian_caplet_factorized}
	\frac{\partial^2 \text{Caplet}_i}{\partial p_u \partial p_v}
	=
	\mathcal{M}'_i(\Pi)\frac{\partial \Sigma_i^2}{\partial p_u}\frac{\partial \Sigma_i^2}{\partial p_v}
	+
	\mathcal{M}_i(\Pi)\frac{\partial^2 \Sigma_i^2}{\partial p_u \partial p_v}.
\end{equation}

\paragraph{Term 2: Second Derivatives of Variance}
The second term requires the Hessian of the variance function $\Sigma_i^2$. Since $\Sigma_i^2$ is smooth, Schwarz's theorem applies, i.e., $\frac{\partial^2}{\partial p_u \partial p_v} = \frac{\partial^2}{\partial p_v \partial p_u}$. The 15 unique explicit formulas for these second-order derivatives are provided in Appendix~\ref{app:second_order_derivatives}.

\subsection{Statistical Diagnostics Tools}
\label{subsec:diagnostic_tools}

Based on the WLS equivalence and the analytical derivatives, we define the core tools for our analysis. We provide the diagnostic tools in matrix-algebraic form for easy implementation. However, they are formally derived from the theory of robust statistics, where estimators are viewed as functionals mapping probability measures into the parameter space (see Appendix~\ref{app:theoretical_foundations}). 
The Influence Function defined below is the finite-sample empirical analogue of the corresponding asymptotic influence-based linearization.

\subsubsection{The Weighted Hat Matrix}

Let $\mathbf{W} = \operatorname{diag}(w_1, \dots, w_m)$ be the diagonal matrix of weights. To ensure a symmetric, true orthogonal projector, we define the Weighted Hat Matrix $\mathbf{H}_w \in \mathbb{R}^{m \times m}$ as
\begin{equation}
	\label{eq:hat_matrix}
	\mathbf{H}_w = \mathbf{W}^{1/2}\mathbf{J}(\mathbf{J}^\top \mathbf{W}\mathbf{J})^{-1}\mathbf{J}^\top \mathbf{W}^{1/2},
\end{equation}
where the inverse is understood as the \emph{Moore--Penrose pseudo-inverse} $(\mathbf{J}^\top\mathbf{W}\mathbf{J})^{+}$ whenever $\mathbf{J}^\top \mathbf{W}\mathbf{J}$ is singular; see, e.g., \cite{BenIsraelGreville2003}. As a global convention for this paper, any standard matrix inverse appearing in diagnostic definitions generalizes to this pseudo-inverse in singular cases. Because $\mathbf{W}$ is positive definite, $\mathbf{H}_w$ is the orthogonal projector onto the column space of $\mathbf{W}^{1/2}\mathbf{J}$. In particular, its trace is equal to the rank of the Jacobian:
\[
\operatorname{trace}(\mathbf{H}_w) = \operatorname{rank}(\mathbf{W}^{1/2}\mathbf{J}) = \operatorname{rank}(\mathbf{J}).
\]
If the columns of $\mathbf{J}$ are linearly independent (full column rank $p$), then $\operatorname{trace}(\mathbf{H}_w)=p$. The diagonal elements $h_{kk} = (\mathbf{H}_w)_{kk}$ are the \textit{Leverage Scores}. Notably, these diagonals are algebraically identical to those of the oblique projector $\mathbf{J}(\mathbf{J}^\top \mathbf{W}\mathbf{J})^{-1}\mathbf{J}^\top\mathbf{W}$ mapping back to the original ambient space. A Leverage Score close to 1 indicates that the model is forced to fit a specific data point almost exactly. In the context of asymptotic statistics, such points signal a potential violation of the Noether condition, where the parameter estimate becomes overly dependent on a single observation (see Appendix~\ref{app:geometric_interpretation}).

As stated above and to ensure numerical stability when the Jacobian matrix becomes rank-deficient, the matrix inversion in Eq.~\eqref{eq:hat_matrix} is implemented using the Moore--Penrose pseudo-inverse via Singular Value Decomposition (SVD). Unlike Tikhonov (ridge) regularization, which replaces the projection by a continuously regularized approximation leading to non-integer effective degrees of freedom, the SVD-based approach truncates singular values below a prescribed tolerance, preserving the idempotence of the projection matrix $\mathbf{H}_w$ and therefore keeps \(\operatorname{tr}(\mathbf{H}_w)\) on exact integer levels. This is important
for interpretation: the Effective Degrees of Freedom are meant to reflect true rank loss and active parameter constraints, not a regularization artifact.

\subsubsection{Parameter Uncertainty and Correlation}\label{subsec:uncertainty}
To quantify local parameter uncertainty, we rely on the \textit{Fisher Information Matrix (FIM)} in the local non-linear WLS/asymptotic calibration framework. Conditional on the observed cap portfolio and the realized weight matrix $\mathbf{W}$, the FIM approximates the curvature of the objective function landscape at $\hat{\Pi}$. Analytically, it is defined as
\begin{equation}
	\mathcal{I}(\hat{\Pi}) \approx \frac{1}{\sigma^2}\mathbf{J}^\top \mathbf{W}\mathbf{J}.
\end{equation}
The FIM provides a local measure of parameter identifiability: steep curvature indicates that the corresponding parameter directions are well determined by the observed design and realized weights. Conversely, low Fisher Information (or a high condition number) indicates a flat objective surface, where small perturbations in market prices can lead to large shifts in the calibrated values.

Under the usual regularity conditions for this local WLS approximation (see Appendix~\ref{app:theoretical_foundations}), the estimator $\hat{\Pi}$ is treated as asymptotically normal. Its covariance matrix is approximated by the inverse, or in singular cases the Moore--Penrose pseudo-inverse, of the Fisher Information Matrix:
\begin{equation}
	\label{eq:cov}
	\widehat{\operatorname{Cov}}(\hat{\Pi}) \approx \mathcal{I}(\hat{\Pi})^{-1} = \sigma^2 (\mathbf{J}^\top \mathbf{W}\mathbf{J})^{-1},
\end{equation}
where $\sigma^2$ is the estimated variance of the residuals.

The covariance matrix in Eq.~\eqref{eq:cov} depends on the estimated scale of the residuals, $\sigma^2$. Traditionally, this scale is estimated using the \textit{Mean Squared Error (MSE)}:
\begin{equation}
	\hat{\sigma}^2_{\text{MSE}} = \frac{1}{m-p} \sum_{k=1}^m w_k \left( y^{\text{Mkt}}_k - g_k(\hat{\Pi}) \right)^2.
\end{equation}
However, the MSE is highly sensitive to outliers. As our empirical analysis in Section~\ref{sec:empirical_analysis} reveals, the short-end instruments (e.g., the 3Y cap) often exhibit disproportionately large relative pricing errors. Since the MSE squares these deviations, a single poorly fitted observation can drastically inflate the global variance estimate, leading to artificially wide confidence intervals for all parameters. This phenomenon is known as the \textit{masking effect}.

To obtain a more reliable measure of the typical noise level in the market data, we employ the \textit{Median Absolute Deviation (MAD)} as a robust scale estimator for $\sigma$. Let
\[
e_k = \sqrt{w_k}\bigl(y^{\text{Mkt}}_k - g_k(\hat{\Pi})\bigr),
\qquad k=1,\dots,m,
\]
denote the weighted residuals. The MAD estimator is then defined as
\begin{equation}
	\hat{\sigma}_{\text{MAD}}
	=
	1.4826 \cdot \operatorname{median}_{1 \leq k \leq m}
	\left(
	\left| e_k - \operatorname{median}_{1 \leq j \leq m}(e_j) \right|
	\right).
\end{equation}
Since $w_k = (y_k^{\text{Mkt}})^{-2}$, these weighted residuals coincide with the relative pricing errors. The constant $1.4826 \approx 1/\Phi^{-1}(0.75)$ ensures consistency under normality. Because it is based on median absolute deviations rather than squared residuals, the MAD is much less sensitive to a small number of extreme boundary residuals and therefore provides a more stable estimate of the typical local noise level.

\paragraph{Handling Parameter Constraints via Transformations and the Delta Method}
A direct application of the covariance estimator $\widehat{\operatorname{Cov}}(\hat{\Pi})$ relies on an asymptotic normal approximation in an unconstrained Euclidean space. However, the parameters of the G2++ model are subject to natural physical constraints, such as $|\rho_{xy}| \leq 1$ and $a_x, a_y, \sigma_x, \sigma_y > 0$. Consequently, standard confidence intervals constructed directly in the original parameter space may violate these admissibility conditions (e.g., implying a correlation parameter $\rho_{xy}$ of $-1.05$).

To address this, we apply a componentwise transformation $\eta: \mathcal{D} \subset \mathbb{R}^5 \to \mathbb{R}^5$ that maps the constrained parameter domain $\mathcal{D}$ bijectively onto $\mathbb{R}^5$. The choice of transformation is guided by two complementary principles: enforcing the natural parameter constraints via smooth bijections and improving the local distributional properties through \textit{Variance-Stabilizing Transformations} (VST).

For interior correlation values $|\rho_{xy}| < 1$, we employ the Fisher $z$-transformation $\eta(\rho) = \operatorname{arctanh}(\rho)$. This mapping arises as the solution to the variance-stabilizing differential equation where the asymptotic variance scales proportionally to $(1-\rho^2)^2$. For strictly positive parameters, such as the mean-reversion speeds and volatilities, we apply the logarithmic transformation $\eta(p) = \log(p)$, which is variance-stabilizing when the variance scales with the squared parameter magnitude (see Appendix~\ref{app:vst_theory}).

For the mean-reversion and volatility parameters, the logarithmic scale ensures that variations are measured in relative rather than absolute terms, naturally enforcing positivity. Similarly, the $\operatorname{arctanh}(\rho_{xy})$ mapping translates the bounded correlation range to the entire real line, allowing for symmetric perturbations around the estimate.

Under the local regularity assumptions required for the Delta Method, in particular differentiability of the transformation and local nonsingularity away from active boundaries, the transformed estimator $\eta(\hat{\Pi})$ is asymptotically normal. Its covariance matrix is obtained via the multivariate Delta Method:
\begin{equation}
	\operatorname{Cov}(\eta(\hat{\Pi})) \approx \mathbf{J}_{\eta}(\hat{\Pi}) \cdot \widehat{\operatorname{Cov}}(\hat{\Pi}) \cdot \mathbf{J}_{\eta}(\hat{\Pi})^\top,
\end{equation}
where $\mathbf{J}_{\eta}(\hat{\Pi}) \in \mathbb{R}^{5 \times 5}$ denotes the diagonal Jacobian matrix of the transformation $\eta$. Confidence intervals are first constructed in this unconstrained space and subsequently mapped back to the original domain via the inverse transformation $\eta^{-1}$. This procedure yields boundary-respecting confidence intervals that are better aligned with the local curvature of the statistical model. Near active boundaries, these intervals should be interpreted diagnostically rather than as uniform asymptotic guarantees.

\subsubsection{The Influence Function}
The \textit{Influence Function (IF)} measures the local impact of an observation on the estimated parameters within the WLS formulation, conditional on the realized weight matrix $\mathbf{W}$. The influence vector for the $k$-th observation is:
\begin{equation}
	\label{eq:influence_function}
	\operatorname{IF}(y_k; \hat{\Pi})
	=
	\underbrace{\left(y_k^{\text{Mkt}} - g_k(\hat{\Pi})\right)}_{\text{Residual}}
	\cdot
	\underbrace{\left[ (\mathbf{J}^\top\mathbf{W}\mathbf{J})^{-1} (\mathbf{J}_k)^\top w_k \right]}_{\text{Sensitivity}},
\end{equation}
where $\mathbf{J}_k$ denotes the $k$-th row of the Jacobian.  
We summarize this into a scalar \textit{Influence Score} using the Euclidean norm $\|\operatorname{IF}(y_k)\|$. The IF is defined up to
a multiplicative constant; since we are interested only in relative rankings across
instruments, this scaling does not affect the comparison.

\section{Implementation and Study Design}
\label{sec:implementation}

\subsection{Data Description and Processing}
The empirical analysis is performed on a historical time series of Euro At-The-Money (ATM) interest rate caps. The sample spans from September 1, 2016, to June 2, 2025, and contains 2157 trading days. This period covers distinct market regimes, including the negative interest rate environment, the post-pandemic recovery, and the rapid increase in interest rates during 2022--2023.

\paragraph{Cap Data}
The dataset, obtained from \textit{LSEG Workspace}, consists of daily quotes for ATM caps with a 6-month tenor. The final calibration portfolio contains caps with maturities
\[
T \in \{3,4,5,6,7,8,9,10,12,15,20,25,30\}\ \text{years}.
\]
For each trading day and maturity, the dataset includes the ATM strike and the corresponding market price.

\paragraph{Term Structure and Discounting}
To ensure consistency with the theoretical framework of the G2++ model, we adopt a single-curve framework throughout the empirical analysis. The underlying risk-free term structure is proxied by the Euro Overnight Index Swap (OIS) curve, based on market data linked to the Euro Short-Term Rate (ESTR) (or EONIA prior to the transition).
This choice is a deliberate simplification: for the ALM and ESG applications considered here, the single-curve approach remains a common setup. The analysis therefore studies the local calibration geometry under a consistent risk-free curve convention rather than reproducing a multi-curve trading framework.
The OIS maturity grid used in the implementation contains 55 maturity points and ranges from approximately one day to 50 years. At observed market maturities, discount factors are taken directly from the OIS discount data. For intermediate maturities not directly observed in the data, discount factors are obtained via Svensson interpolation. 

The resulting discount curve serves two critical roles in the calibration: it determines the initial discount factors $P(0,T)$ used for valuation, and it acts as the input for the deterministic function $\psi(t)$ (see Eq.~\eqref{eq:psi}). This ensures that the G2++ model perfectly fits the initial term structure at time $t=0$.

Only dates for which complete cap prices, cap strikes, and OIS curve data are available are kept. Trading days with incomplete information were filtered out of the time series.

\subsection{Efficient Jacobian Implementation}
The analysis is implemented in R. Instead of repeatedly approximating the Jacobian matrix $\mathbf{J}$ by finite differences in the daily diagnostic pipeline, we employ the \textit{Jacobian Factorization} derived in Section~\ref{subsec:jacobian_derivation}. This avoids additional numerical approximation error while reducing computational cost.

An important feature of ATM cap calibration is that each cap has its own strike rate $K_k$, so that the price-to-variance sensitivity factor $\mathcal{M}_i$ varies across instruments. Our implementation assembles the Jacobian row-by-row as
\begin{equation}
	\mathbf{j}_k = \mathbf{c}_k \cdot \mathbf{M}_{\mathrm{diag}}(K_k) \cdot \mathbf{S}_{\mathrm{deriv}},
\end{equation}
where $\mathbf{c}_k$ is the composition vector identifying the caplets contributing to cap $k$, $\mathbf{M}_{\mathrm{diag}}(K_k)$ is the diagonal matrix of $\mathcal{M}_i$ factors, and $\mathbf{S}_{\mathrm{deriv}}$ is the matrix of variance derivatives reused across all instruments. This vectorized implementation avoids repeated finite-difference evaluations in the daily diagnostic pipeline.

\subsection{Experimental Procedure}
For each day \(t\) in the historical dataset, the following procedure is executed:
\begin{enumerate}
	\item \textbf{Calibration:} The G2++ model parameters \(\hat{\Pi}_t\) are estimated by minimizing the RMSRE objective function. The parameter bounds are chosen as
	\[
	\sigma_x,\sigma_y \in [10^{-5},1], \qquad
	a_x,a_y \in [10^{-5},10], \qquad
	\rho_{xy} \in [-1,1].
	\]
	The baseline optimization is performed with the derivative-free solver BOBYQA, using a relative tolerance of \(10^{-8}\) and a maximum of 20{,}000 function evaluations.
	
	\item \textbf{Linearization:} Using \(\hat{\Pi}_t\), the analytical Jacobian \(\mathbf{J}_t\) and the weight matrix \(\mathbf{W}_t\) are constructed.
	
	\item \textbf{Diagnostics:} The key diagnostic matrices, specifically the Weighted Hat Matrix \(\mathbf{H}_{w,t}\) and the Fisher Information Matrix, are computed in order to derive leverage scores, effective degrees of freedom, parameter uncertainty measures, and influence diagnostics.
\end{enumerate}

\paragraph{Robustness of the calibration design.}
To assess sensitivity to the numerical setup, we ran additional checks beyond the baseline calibration. These included alternative starting values, a comparison of BOBYQA with the derivative-free Nelder--Mead algorithm, and reduced calibration portfolios in which selected boundary maturities were removed, in particular the 3Y cap and, separately, the 25Y and 30Y caps. The resulting diagnostics were qualitatively stable across these variants: in particular, the boundary concentration of leverage and the main EDoF patterns remained visible. Therefore, the baseline specification reported should be understood as a representative implementation rather than a unique numerical configuration under which the empirical findings arise.

\section{Empirical Analysis: The Euro Cap Market (2016--2025)}
\label{sec:empirical_analysis}

In this section, we apply the diagnostic framework to the historical Euro ATM cap dataset. Our objective is to answer the four diagnostic questions posed in the introduction: How good is the fit in price space? Which maturities drive the calibration? When does the model lose effective dimensionality? And how large is the remaining parameter uncertainty even when the price fit appears satisfactory?

The empirical analysis reveals four features of the calibration problem: time-varying goodness of fit in price space, boundary-dominated calibration geometry, repeated losses of effective dimensionality due to active parameter constraints, and strongly time-varying parameter uncertainty. Taken together, these diagnostics point to a regime shift in local parameter stability around 2022.

\subsection{Goodness of Fit over Time}
\label{subsec:rmsre_time}

\begin{figure}[htbp]
	\centering
	\includegraphics[width=\textwidth]{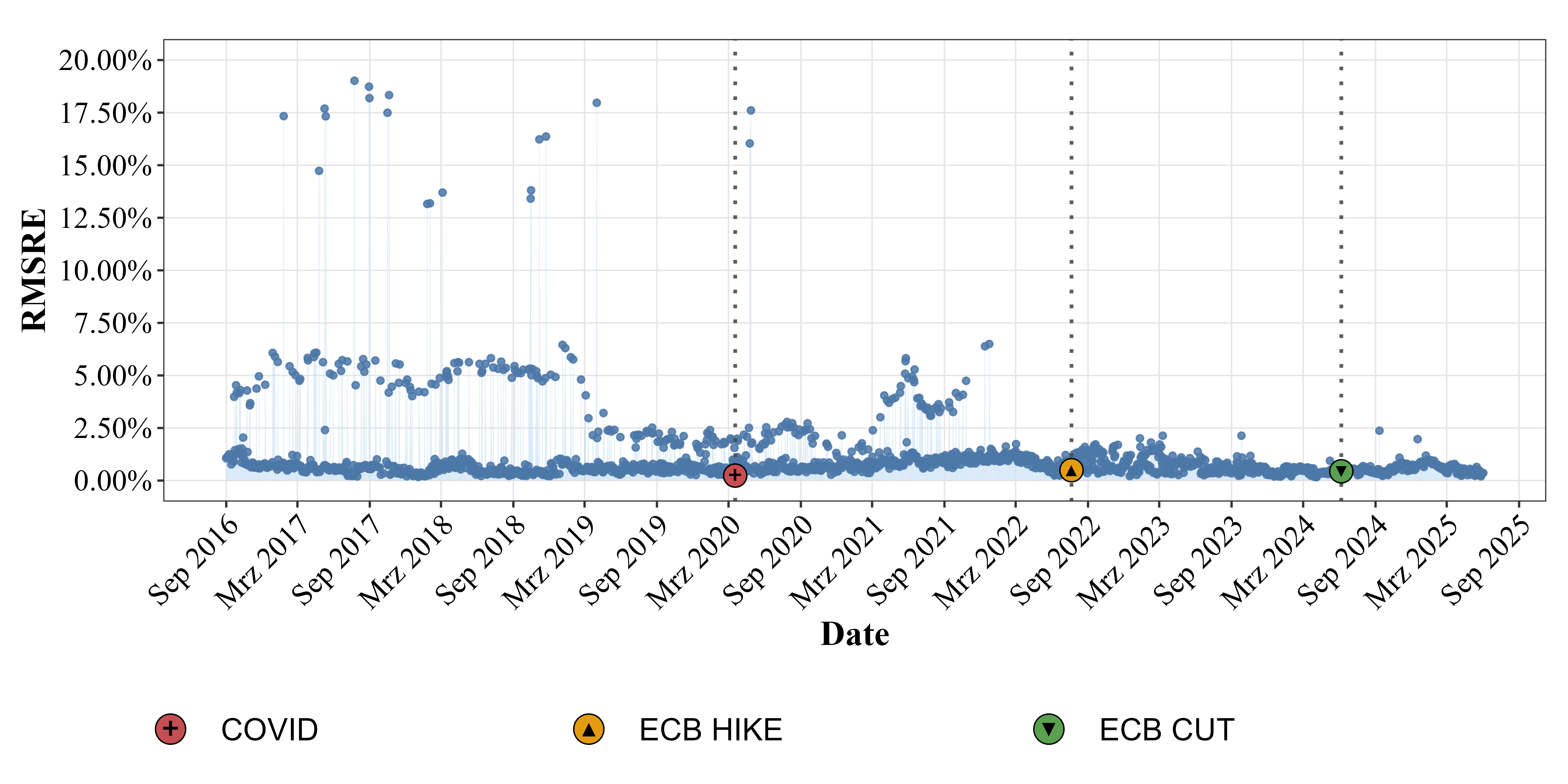}
	\caption{\textbf{Daily goodness of fit of the G2++ cap calibration.}
		The figure shows the daily Root Mean Squared Relative Error (RMSRE) over the full sample 2016--2025.
		Fit quality is clearly time-varying: the early sample contains extended periods of elevated RMSRE and several extreme outliers, whereas from 2022 onward the RMSRE is low on most days. The three event markers indicate the reference dates used throughout the empirical analysis. From the RMSRE perspective alone, these dates do not appear as exceptional calibration failures. Thus, for these examples, economically important market dates do not necessarily appear as poor-fit days.}
	\label{fig:rmsre_time}
\end{figure}

Before turning to leverage, influence and identifiability, it is useful to distinguish between two notions that are often conflated in practice:
goodness of fit in price space and stability in parameter space. Figure~\ref{fig:rmsre_time} reports the daily RMSRE of the calibrated G2++ model over the full sample 2016--2025. The vertical dotted lines mark the three temporal landmarks introduced above: the COVID shock in March 2020, the first ECB rate hike in July 2022, and the first ECB rate cut in June 2024.

Overall, the calibration achieves a low RMSRE on many days throughout the sample, often around \(1\%\). At the same time, fit quality is clearly time-varying rather than uniform. In the earlier part of the dataset, from 2016 to 2021, the RMSRE frequently reaches levels around \(4\%\) to \(6\%\), and several isolated days exhibit substantially larger deviations, with extreme spikes close to \(20\%\). From 2022 onward, the daily RMSRE is, for the vast majority of days, comparatively low and often remains below roughly \(1\%\) to \(1.5\%\), despite occasional increases.

This time pattern is important for the interpretation of the subsequent diagnostics.
The post-2022 instability identified below in the Effective Degrees of Freedom and in the PCA is therefore not the consequence of a generally deteriorating price fit. Instead, a substantial part of the later sample combines a comparatively good fit to the market prices with visible instability in parameter space.
This is precisely the distinction our framework is designed to make visible: a small RMSRE confirms that model prices lie close to observed market prices, but it does not necessarily imply that the corresponding parameter vector is stably identified or that the local calibration geometry is well behaved. From a practical perspective, Figure~\ref{fig:rmsre_time} therefore serves as a baseline diagnostic.

The three marked dates are not selected because they are outliers in RMSRE. All three lie in periods in which the aggregate fit is relatively good. A purely fit-based assessment would therefore not treat these dates as problematic. This makes them useful reference points for the subsequent analysis: the same dates can be compared across leverage, influence, EDoF, and uncertainty diagnostics, where their differences become more visible.

\subsection{Leverage and Influence Analysis}
\label{subsec:leverage_influence}

\begin{figure}[!htbp]
	\centering
	\begin{subfigure}[b]{0.9\textwidth}
		\includegraphics[width=\textwidth]{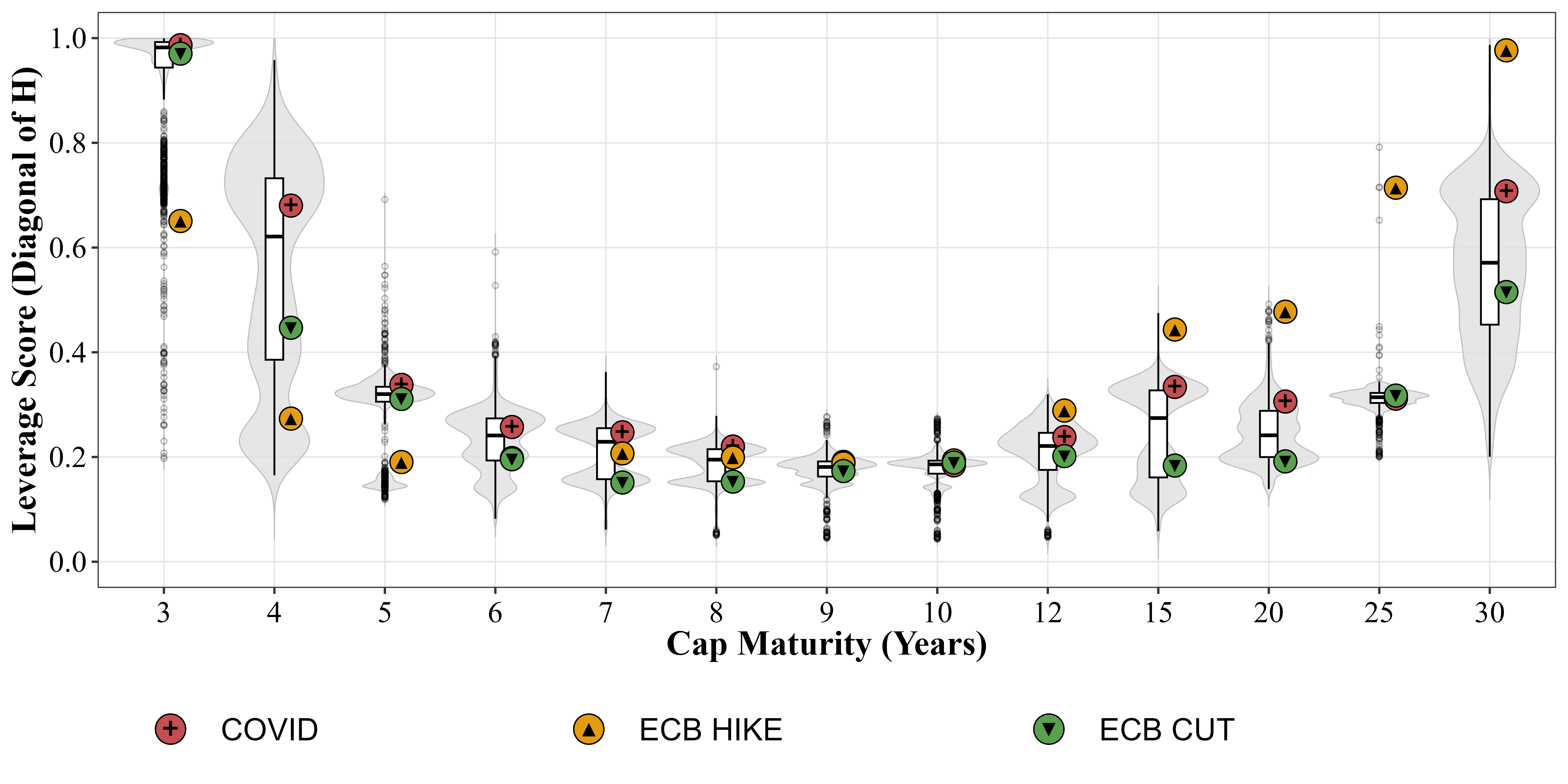}
		\caption{Leverage Scores (Geometry)}
	\end{subfigure}
	\hfill
	\begin{subfigure}[b]{0.9\textwidth}
		\includegraphics[width=\textwidth]{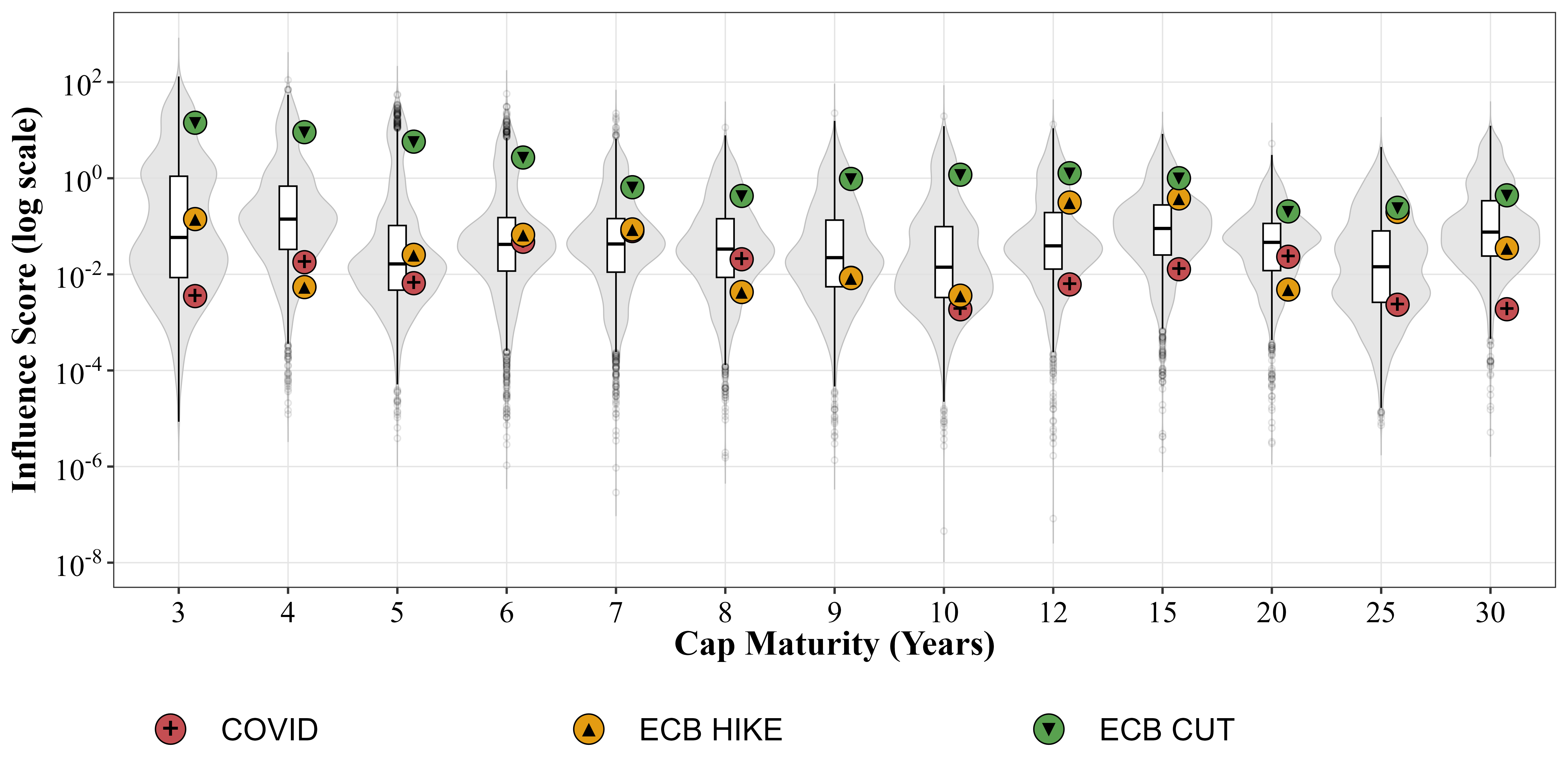}
		\caption{Influence Scores (Sensitivity)}
	\end{subfigure}
	\caption{\textbf{Leverage vs. Influence.}
		The violin layer, following the box plot--density trace construction of Hintze and Nelson~\cite{HintzeNelson1998}, shows the distribution over all calibration days, while the boxplot summarizes its center and spread. The symbols mark the three reference dates used throughout the paper.
		(a) The leverage profile is boundary-dominated and asymmetric. Leverage is concentrated at the maturity boundaries; in the 3Y cap, leverage exceeds \(0.95\) on 73.1\% of all days and exceeds \(0.99\) on 32.7\% of all days.
		(b) The Influence Scores reveal a more diffuse picture on the log scale: no single maturity dominates the Influence Scores across the full sample, while the short end exhibits the most extreme tail events.}
	\label{fig:leverage_influence}
\end{figure}

A key question in model validation is identifying which instruments drive the resulting parameter values. We study this using two complementary metrics: the \textit{Leverage Score} (quantifying potential impact based on model geometry) and the \textit{Influence Score} (quantifying actual impact based on pricing errors). The relationship between these two diagnostics is illustrated in Figure~\ref{fig:leverage_influence}.

These plots should be read as one diagnostic layer within a broader calibration assessment. They do not by themselves prescribe a change of the calibration portfolio or of the weighting scheme. Rather, holding the model, objective function, and market data structure fixed, they show where the local calibration geometry and the realized residual sensitivity are concentrated.

The Leverage Scores (Panel a) are boundary-dominated and asymmetric. Local leverage is concentrated at both the short end (3Y, 4Y) and the long end (30Y) of the maturity spectrum.

Notably, the leverage score for the 3Y cap frequently approaches the upper bound of \(1\). In the Noether-condition terminology this is not proof of asymptotic failure, but it signals that the calibration can become locally dependent on a single market quote. This highlights how strongly the local calibration geometry can concentrate at the maturity boundaries.

However, the Influence Scores (Panel b) reveal that the \textit{realized} impact on the parameters follows a different and substantially more differentiated distribution.

In contrast to the leverage profile, the influence scores do not single out one unique long-end maturity as a dominant driver across the full sample. On the logarithmic scale, several maturities exhibit comparable influence levels, while the upper tails differ more strongly. In particular, the short end shows the most extreme tail events, indicating that short-maturity pricing shocks can translate into large parameter displacements once combined with the very high local leverage.

This suggests the following distinction. The leverage plot identifies the maturities that are important in the local calibration geometry, whereas the influence plot reflects how this geometric importance combines with the realized residuals. Hence, although the boundary-dominated geometry remains clearly visible, realized parameter instability is spread more broadly across maturities. In particular, the short end is the primary source of potential instability, while the long end contributes a more persistent, but less uniquely dominant, pull on the calibrated parameters.

In summary, the influence diagnostic adds an important nuance to the calibration problem. The boundary maturities determine a large part of the local fit geometry, but realized parameter instability is not driven by a single long-end quote. Instead, it arises from the interaction between high leverage at the boundaries and maturity-specific residuals. This instability is often an indicator that the model is hitting its parameter boundaries, which we investigate in the following section.

The three event markers illustrate that low-RMSRE days can still differ substantially in their local diagnostics. The COVID date does not stand out strongly in leverage or influence. The ECB hike is associated with a stronger role of the long end, whereas the ECB cut combines high short-end leverage with the largest influence value among the three reference dates. A joint numerical comparison of these three dates is given in Table~\ref{tab:event_scenarios} below.

\subsection{Effective Degrees of Freedom and Parameter Constraints}
\label{subsec:dof}

\begin{figure}[htbp]
	\centering
	\includegraphics[width=\textwidth]{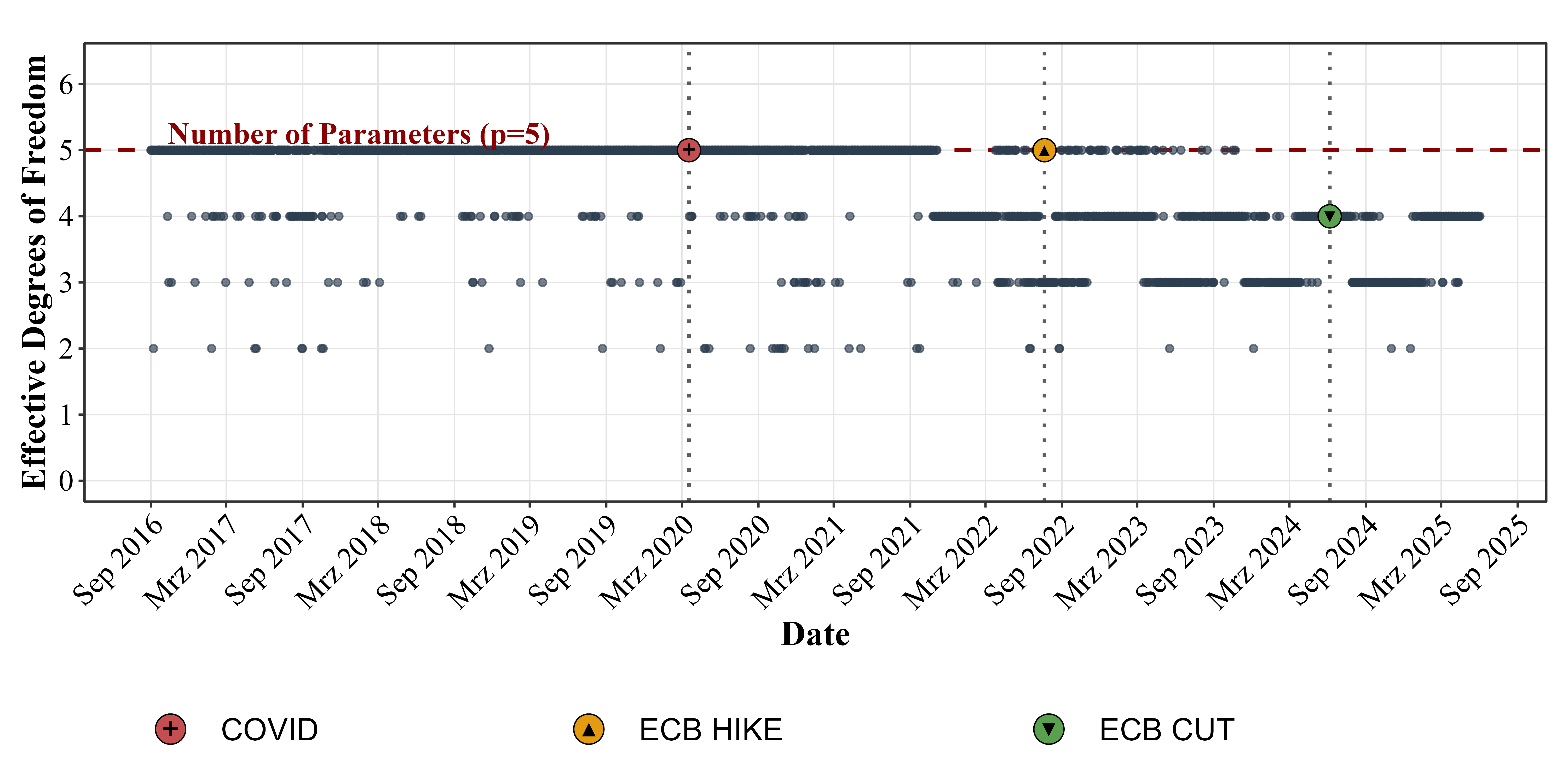}
	\caption{\textbf{Effective Degrees of Freedom (Trace of the Weighted Hat Matrix).} Most daily calibrations lie exactly on the theoretical maximum of $5$, indicating full local rank of the Jacobian. The distinct horizontal bands at lower integer values reveal instances of rank loss, typically associated with active parameter constraints. Notably, the period of market transition (2022--2024) shows more frequent switching between these integer EDoF levels.}
	\label{fig:dof}
\end{figure}

A fundamental property of the Weighted Hat Matrix \( \mathbf{H}_w \) is that its trace (the sum of leverages) equals the number of free parameters \( p \). For the G2++ model, this sum equals $5$ whenever the Jacobian matrix possesses full column rank. However, a drop in this trace -- the \textit{Effective Degrees of Freedom} (EDoF) -- signals a loss of local identifiability. Figure~\ref{fig:dof} tracks this diagnostic over time, revealing systematic intervals in which the effective dimension of the calibration problem is reduced.

Our analysis shows that over the vast majority of the 10-year period, the model is locally well identified with the full $5$ degrees of freedom. However, the presence of sharply defined bands at strict lower integer values confirms that on specific days, the effective dimension of the local calibration problem partially collapses. Because we use the SVD-based pseudo-inverse, these bands occur at exact integers, indicating true rank loss rather than a numerical smoothing artifact. For a comparison with a Tikhonov-regularized alternative, see Appendix~\ref{app:edof_regularization_comparison}.

This phenomenon is often associated with the optimizer hitting parameter boundaries; for instance, when the correlation $\rho_{xy}$ reaches $-1$ or a mean reversion speed $a_x$ reaches its lower or upper limit, that parameter effectively ceases to be a free local direction. The fact that the bands in Figure~\ref{fig:dof} occur at integer values is consistent with the influence of active-set bounds in the optimization process, where the optimizer explicitly restricts the parameter space. In our implementation, the calibration space is given by $0 < a_x, a_y \leq 10$, $0 < \sigma_x, \sigma_y \leq 1$ and $|\rho_{xy}|\leq 1$.

In the empirical sample, the most common active constraint is the lower correlation bound: $\rho_{xy}=-1$ occurs on 66.4\% of the trading days in the baseline calibration, and on 69.6\% if a numerical tolerance of $10^{-6}$ is used. The EDoF and leverage plots in the main text use the ordinary full-Jacobian Moore--Penrose construction: all five Jacobian columns are kept, while singular directions are handled by the pseudo-inverse. In this ordinary full-Jacobian diagnostic, boundary dominance also persists on days with \(\rho_{xy}=-1\). If the binding correlation constraint is treated more strictly, however, \(\rho_{xy}\) is no longer counted as a free local parameter. Under this stricter interpretation, boundary dominance remains visible on most, but not all, of the \(\rho_{xy}=-1\) days. Thus, the additional constraint-based diagnostic supports the boundary-concentration interpretation, but should not be read as a universal statement for every constrained calibration day.

The transition around 2022 is particularly informative. While the earlier history shows relatively clean and stable periods at full rank, interrupted by discrete drops to lower integer levels, the period following the 2022 shift exhibits much more frequent switching between these levels. This suggests that during the transition from the negative-rate to the high-rate regime, local identifiability became more fragile and the optimizer interacted more often with the parameter bounds.

This diagnostic highlights an important risk: even if two daily calibrations show similar pricing errors (RMSRE), a drop in EDoF indicates that the resulting parameter set is not equally well determined by the market data. Our framework allows practitioners to detect these states of reduced local identifiability, helping to distinguish between well-identified calibrations and solutions that are only weakly identified by the observed market prices.

The three marked event dates provide a compact contrast set for the diagnostic analysis. Table~\ref{tab:event_scenarios} shows that all three days have RMSRE below \(0.5\%\), but differ substantially in their local geometry. The COVID date has excellent fit and full EDoF, the ECB hike is associated with high long-end leverage, and the ECB cut combines low aggregate RMSRE with rank loss and a large short-end influence value.

\begin{table}[htbp]
	\centering
	\footnotesize
	\caption{\textbf{Three calibration reference cases.} The three dates are marked consistently in the time-series and violin-boxplots. Although all three have low aggregate RMSRE, the local diagnostics lead to different interpretations.}
	\label{tab:event_scenarios}
	\begin{tabular}{llccccp{4.3cm}}
		\toprule
		Date & Event & RMSRE & EDoF & Max leverage & Max influence & Diagnostic reading \\
		\midrule
				2020-03-18 & COVID & 0.24\% & 5.00 & 0.99 (3Y) & 0.08 (7Y) & Full EDoF; persistent 3Y leverage appears structural. \\
		2022-07-21 & ECB hike & 0.49\% & 5.00 & 0.98 (30Y) & 0.38 (15Y) & Good fit with long-end geometry dominance. \\
				2024-06-06 & ECB cut & 0.44\% & 4.00 & 0.97 (3Y) & 14.27 (3Y) & Reduced EDoF with a large short-end influence value. \\
		\bottomrule
	\end{tabular}
\end{table}

The last row illustrates most clearly why a fit-only reading can be misleading: on 2024-06-06 the aggregate RMSRE remains small, although the largest single relative residual reaches 47.9\% at the 3Y cap. Thus, the three dates are useful not because they are RMSRE outliers, but because they show how similar aggregate fit quality can correspond to different local diagnostic states.

\subsection{Regime-Dependent Calibration Geometry}
We next analyze the stability of the solution space using Principal Component Analysis (PCA). Projecting the 5-dimensional parameter vectors onto the first two principal components visualizes the movement of the calibration over time and identifies distinct market regimes.

\begin{figure}[htbp]
	\centering
	\includegraphics[width=0.85\textwidth]{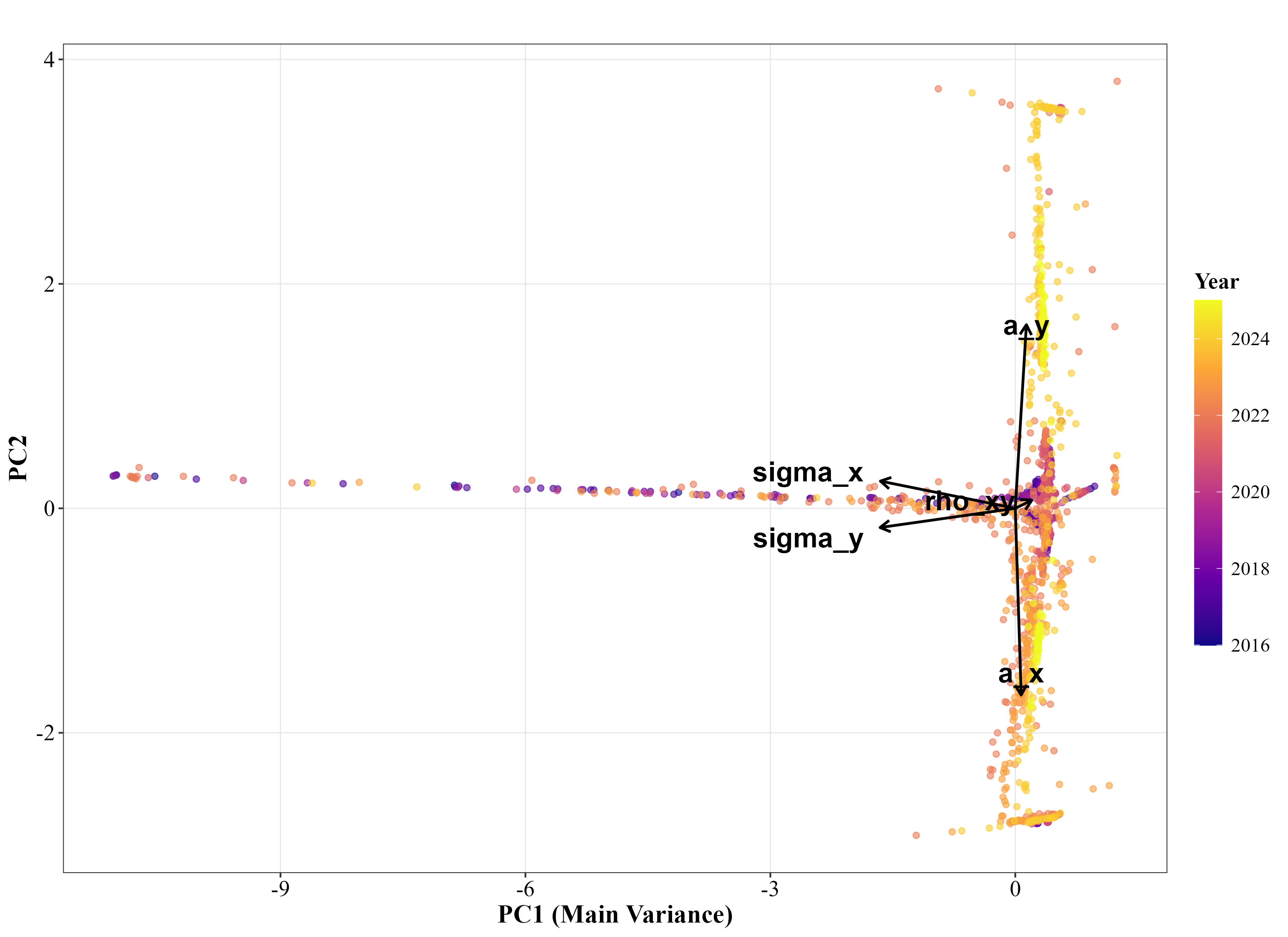}
	\caption{\textbf{PCA Biplot of Calibration Parameters.} The points represent daily calibrations colored
		by year. The vectors (arrows) indicate the loading of the original parameters on the principal components. The horizontal PC1 captures market volatility ($\sigma_{x}$, $\sigma_{y}$), while the vertical PC2 captures structural parameters such as mean reversion and correlation.}
	\label{fig:biplot}
\end{figure}

Figure~\ref{fig:biplot} provides the interpretation of the axes. The horizontal axis (PC1) captures the variation in the volatility parameters $\sigma_x$ and $\sigma_y$.  Vertical movement (PC2) is driven by changes in mean reversion speeds ($a_x, a_y$) and the correlation coefficient ($\rho_{xy}$). Notably, the mean reversion parameters $a_x$ and $a_y$ point in nearly opposite directions along the vertical axis, suggesting a functional trade-off where the optimizer adjusts one speed to compensate for changes in the other.

\begin{figure}[htbp]
	\centering
	\includegraphics[width=0.85\textwidth]{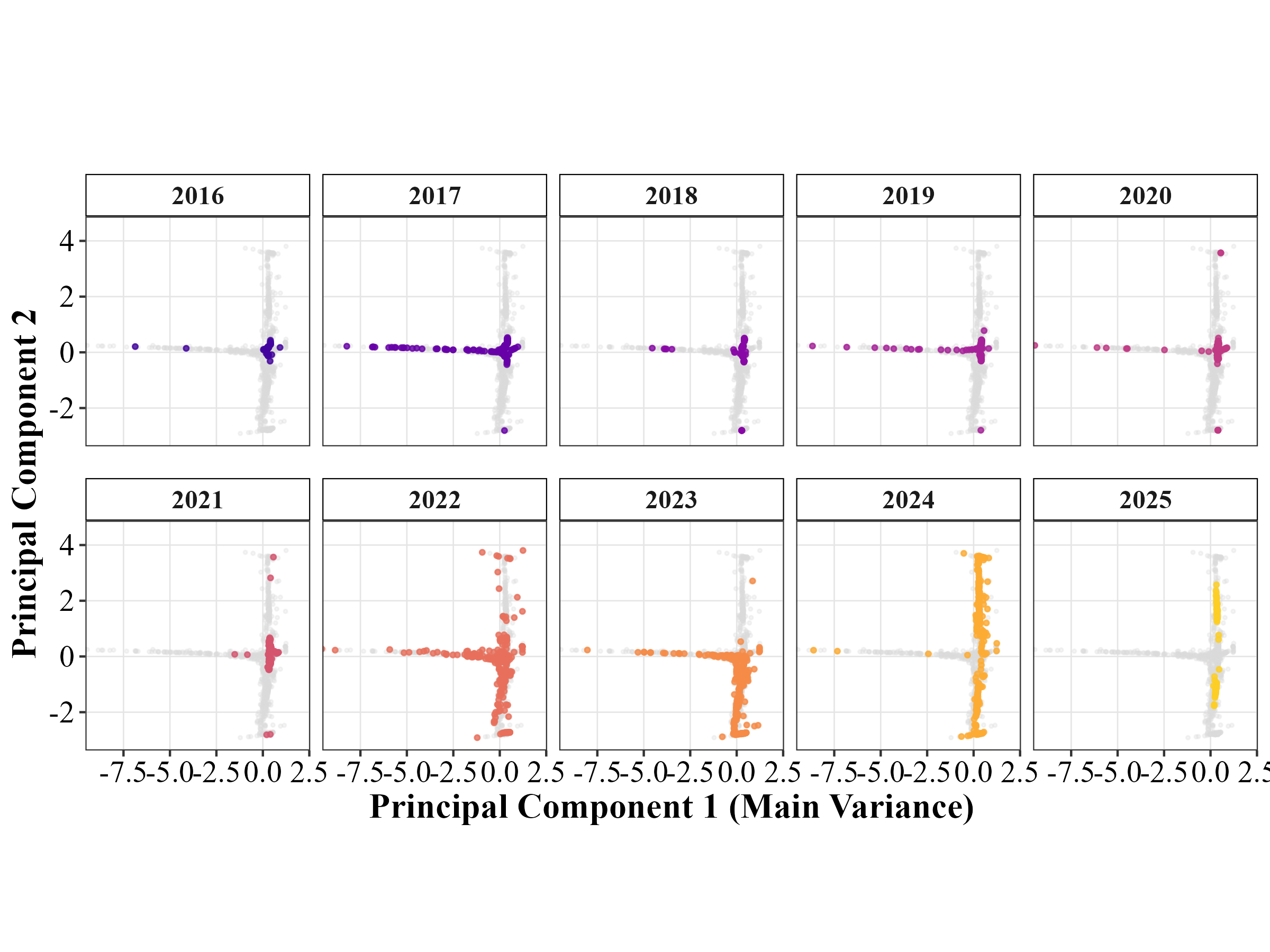}
	\caption{\textbf{Temporal Evolution of the Calibration Regime.} The faceted view highlights the migration from the low-volatility regime (2016--2021) to the high-volatility regime (2022--2025). The grey background illustrates all points of the calibration history.}
	\label{fig:pca_faceted}
\end{figure}

As shown in the faceted view in Figure~\ref{fig:pca_faceted}, the calibration history reveals a distinct pattern of structural stability, followed by a period of pronounced vertical expansion, which weakens by the end of the dataset in 2025. During the \textit{Horizontal Regime} (2016--2021), the calibrations are confined to a narrow, horizontal band at the bottom of the PCA space. This indicates that the parameters associated with PC2 (primarily mean reversion and correlation) remained almost constant during the negative interest rate environment, with nearly all variation occurring along the volatility axis.

This stability changed during the \textit{Transition Phase} (2022--2024). Starting in 2022, the model diverges from its horizontal confinement, exhibiting significant up- and downward movement in PC2. This vertical expansion indicates that the structural parameters, which were previously relatively stable, began to fluctuate more strongly. This period of expansion in the PCA space closely aligns with the increased switching observed in the Effective Degrees of Freedom (Figure~\ref{fig:dof}), marking a phase in which the calibration more frequently moved between interior and boundary solutions.

The 2025 data consolidates into a vertical cluster, where the amplitude of the daily variation along PC2 has significantly decreased compared to the 2022--2024 phase. This indicates a marked shift in the calibration mechanics: in the current market environment, the volatility scale is relatively stable, while the daily fluctuations are now primarily driven by adjustments to the mean reversion and correlation parameters.

This visualization suggests that the instability observed in the parameters is a reflection of the changing information content in the market data. The transition between these regimes corresponds to the period in which the Effective Degrees of Freedom (Figure~\ref{fig:dof}) switches most frequently between integer levels, indicating that the parameter vector became less stably identified while the calibration moved into a different region of parameter space.

The three event dates also help to locate the different parts of the PCA trajectory. The COVID date lies close to the pre-2022 horizontal corridor. The ECB hike falls into the transition period, where the trajectory starts to expand vertically. The ECB cut belongs to the later post-2022 cluster, in which the calibration remains boundary-sensitive despite low RMSRE. Thus, the events should not be interpreted as identical stress cases. Rather, they provide common reference points for comparing fit, local geometry, and parameter movement.

\subsection{Parameter Uncertainty and Confidence Intervals}
\label{subsec:ci_ay}

The previous subsections focused on geometric diagnostics (leverage, influence, and effective degrees of freedom) and on the movement of the point estimates $\hat\Pi_t$ over time. We now explicitly quantify \textit{parameter uncertainty} by constructing confidence intervals (CIs) for the calibrated parameters based on the Fisher Information Matrix and the Variance Stabilizing Transformations introduced in Section~\ref{subsec:uncertainty}.

These intervals are therefore local diagnostic intervals, conditional on the realized weights and on the local WLS linearization. Near active constraints or rank-deficient cases, they should be interpreted together with the EDoF and leverage diagnostics rather than as unconditional probabilistic guarantees.

We obtain the covariance matrix of the WLS estimator via \eqref{eq:cov}, interpreting it as a local asymptotic approximation to the sampling variability of $\hat\Pi_t$ under the non-linear regression model. To respect the admissible parameter region ($0 < a_x, a_y \leq 10$, $0 < \sigma_x, \sigma_y \leq 1$ and $|\rho_{xy}|\leq 1$), we apply the log-transformation to $a_x$, $a_y$, $\sigma_x$ and $\sigma_y$, while the Fisher $z$-transformation is used for the correlation. Uncertainty is propagated in the transformed space using the Functional Delta Method and subsequently mapped back to the original scale. By construction, the resulting confidence intervals remain within the admissible parameter region and are generally asymmetric.

\begin{figure}[htbp]
	\centering
	\includegraphics[width=\textwidth]{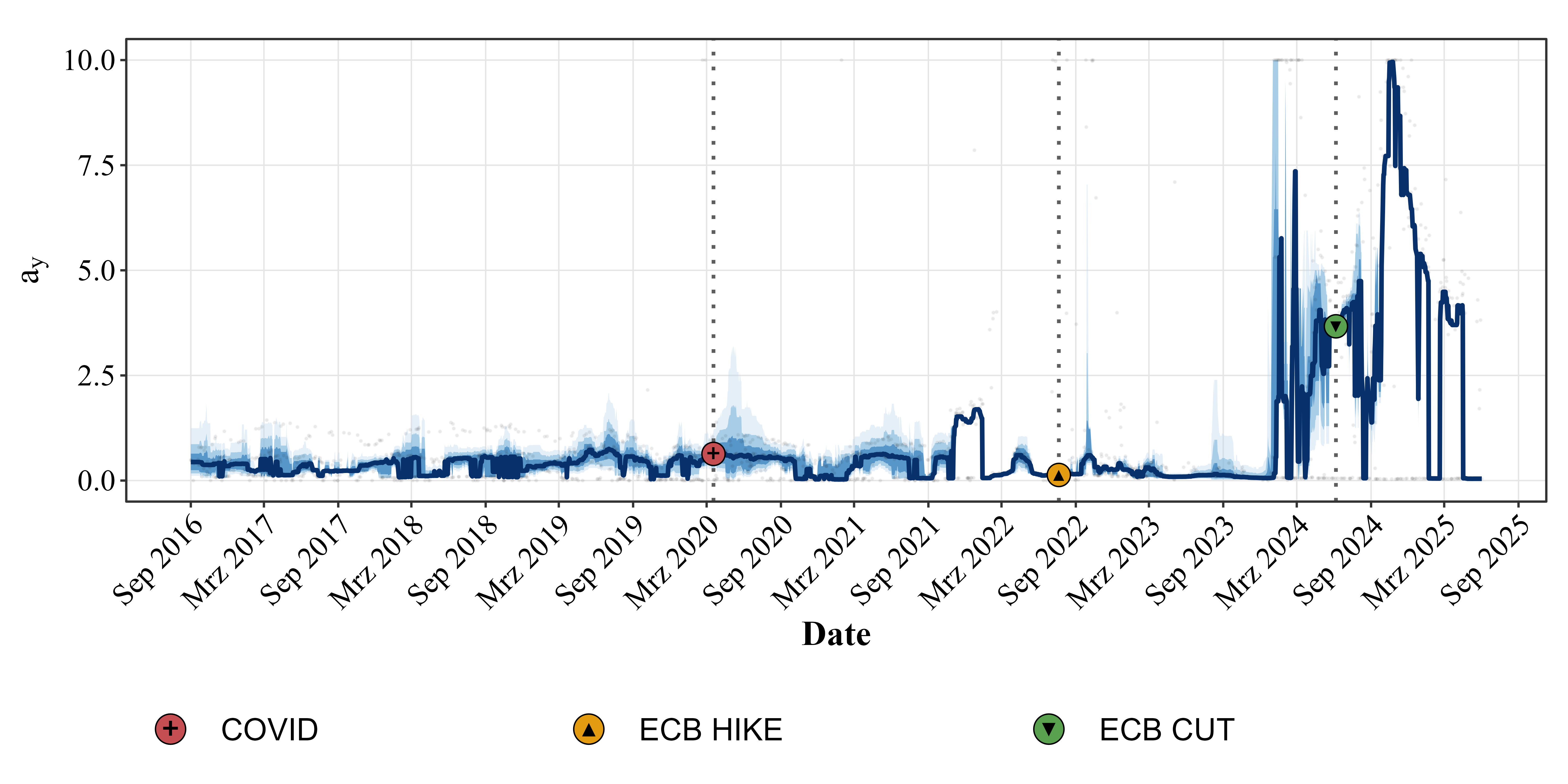}
	\caption{Time-varying confidence intervals for $a_y$. The solid line shows the daily calibrated value $\hat a_{y,t}$, while the shaded areas represent the $50\%$, $75\%$, and $95\%$ confidence intervals obtained via the Variance Stabilizing Transformations, the Functional Delta Method, and robust MAD-based covariance scaling.}
	\label{fig:ci_fan_ay}
\end{figure}

Figure~\ref{fig:ci_fan_ay} illustrates the resulting \textit{fan chart} for the mean reversion speed of the second factor, \(a_y\), using the \emph{robust MAD-based} residual scale.\footnote{Here, MAD refers to the consistency-corrected median absolute deviation, i.e. the version returned by \texttt{mad()} in \textsf{R} with its default factor \(1/\Phi^{-1}(0.75)\approx 1.4826\), which is consistent for the Gaussian scale parameter \(\sigma\). It is therefore not the unscaled raw median absolute deviation.} The dark line shows the daily point estimates \(\hat a_{y,t}\). The shaded areas represent the \(50\%\), \(75\%\), and \(95\%\) CIs obtained from the VST-based intervals. For comparison with the classical MSE-based scaling, see Appendix~\ref{app:classical_vs_robust_ci}.

Several features are noteworthy. First, for extended periods the point estimate \(\hat a_{y,t}\) lies close to the lower part of the admissible region, and the confidence intervals are visibly asymmetric. This confirms that a naive Gaussian approximation on the original scale would frequently be inappropriate near the boundary.

Second, the robust fan chart does \emph{not} exhibit the persistent full-range uncertainty bands that arise under the classical MSE-based scaling in Appendix~\ref{app:classical_vs_robust_ci}. Instead, the broad uncertainty episodes become much more localized. In particular, the periods of increased uncertainty around 2019--2020 remain clearly visible, but the robust scaling avoids interpreting a small number of extreme residuals as evidence of inflated parameter uncertainty. As discussed in Appendix~\ref{app:classical_vs_robust_ci}, this difference reflects the \textit{masking effect} created by MSE-based estimation.

Third, one must interpret episodes in which the confidence bands collapse to the point estimate with care. In the present framework, the covariance matrix is computed using the SVD-based pseudo-inverse $\left(\mathbf{J}^{\top}\mathbf{W}\mathbf{J}\right)^{+}$. When $\mathbf{J}^{\top}\mathbf{W}\mathbf{J}$ loses rank, the pseudo-inverse does not inflate uncertainty in directions that are no longer identified; rather, it truncates these directions in the covariance approximation. As a result, the coordinate-wise variance of \(a_y\) can vanish on some days, so that the corresponding confidence interval collapses to the point estimate. This should \emph{not} be interpreted as evidence of unusually high statistical precision. Instead, such episodes indicate that the local linear approximation has ceased to treat \(a_y\) as a genuinely free direction. Collapsed bands are therefore used as diagnostic geometry indicators and must be interpreted jointly with the Effective Degrees of Freedom in Figure~\ref{fig:dof}, not in isolation.

For the confidence intervals, the event dates mainly serve as common reference points rather than as the dates with the largest uncertainty bands.

Accordingly, the daily calibration \(\hat\Pi_t\) should not be treated as precise merely because the RMSRE is low or because the confidence interval of one coordinate collapses. The more relevant conclusion is that parameter uncertainty is highly time-varying, boundary-sensitive, and closely linked to local identifiability. Such confidence intervals provide a useful measure of parameter risk, but only when interpreted jointly with the geometric diagnostics developed in the previous subsections.

\section{Lessons Learned and Outlook}
\label{sec:conclusion}

\paragraph{Contributions}
This paper develops a statistical framework for the calibration of stochastic interest rate models that links numerical optimization with non-linear regression diagnostics. The industry-standard Root Mean Squared Relative Error (RMSRE) objective is equivalent to a Weighted Least Squares (WLS) problem. This equivalence makes it possible to interpret the calibration problem through the geometry of weighted least squares and to apply established diagnostic tools to the G2++ model.

Our primary technical contribution is the derivation of an exact analytical Jacobian factorization for ATM caps. This enables the computation of diagnostics such as the Weighted Hat Matrix, leverage scores, Influence Functions, and Fisher Information Matrix with improved numerical stability and low additional computational overhead. Furthermore, we addressed the limitations of standard asymptotic theory in constrained parameter spaces by combining the Functional Delta Method with Variance Stabilizing Transformations. This yields confidence intervals that remain within the admissible parameter region and respect the natural boundaries of the G2++ parameter space, such as $|\rho_{xy}| \leq 1$.

The framework should not be read as a replacement for economic or actuarial model validation. It provides a local diagnostic layer for the market calibration at \(t=0\). Its purpose is to identify whether a low calibration error is accompanied by stable local geometry, well-identified parameters, and limited dependence on individual market quotes.

\paragraph{Lessons Learned}
Applying this framework to a decade of Euro ATM cap data yields several insights:

\begin{description}
	\item \textbf{Boundary effects and the asymmetric role of short- and long-end caps.} The leverage and influence diagnostics in Figure~\ref{fig:leverage_influence} show that the local calibration geometry is concentrated at the maturity boundaries rather than in the interior. The leverage profile is boundary-dominated and asymmetric, with particularly high leverage at the short end and elevated leverage at the long end. At the same time, the realized influence is distributed more broadly across maturities than the leverage profile alone would suggest. The short end is the main source of potential instability, because the 3Y cap frequently has leverage close to \(1\), so that short-end anomalies can be amplified into substantial parameter displacements. The long end exerts a more persistent, but less uniquely dominant, pull on the calibrated parameters.
	The three event dates illustrate this distinction. The COVID date is economically important, but remains comparatively quiet in terms of the core diagnostics. The ECB hike date highlights the role of the long end, whereas the ECB cut date is the clearest example of a low-RMSRE calibration in which the short end becomes the dominant source of local instability.
	
	\item \textbf{Regime-dependent identifiability and effective dimensionality.}
	The trace of the Weighted Hat Matrix in Figure~\ref{fig:dof} shows that most daily calibrations lie exactly on the theoretical maximum of \(5\), but repeatedly drop to lower integer levels when parameters hit their bounds or when the local Jacobian loses rank. Hence, the Effective Degrees of Freedom provide a direct diagnostic for local dimensionality loss.
	
	In the pre-2022 period, these integer levels are relatively stable and indicate that one or more parameters are effectively determined by constraints rather than by the market data alone. The PCA in Figures~\ref{fig:biplot} and \ref{fig:pca_faceted} complements this picture. Before 2022, the calibration path is largely confined to a narrow horizontal band dominated by the volatility scale, while mean-reversion speeds and correlation remain comparatively stable. The three marked events are therefore useful reference cases because they do not all create the same diagnostic pattern: a widely observed market shock can remain comparatively quiet in the local diagnostics, whereas later low-RMSRE dates can reveal local identifiability problems.
	
	\item \textbf{The 2022 diagnostic regime shift.}
	Around 2022, the behaviour of the calibration changes qualitatively. The PCA trajectory leaves its earlier horizontal corridor and shows stronger vertical fluctuations. This reflects larger daily movements in mean-reversion speeds and correlation. At the same time, the Effective Degrees of Freedom become less stable, with more frequent switching between integer levels. This indicates repeated transitions between interior and boundary solutions.
	
	By 2025, the cloud of points consolidates into a more compact, vertically oriented cluster. The volatility level is then relatively stable, while day-to-day variation is driven mainly by the mean-reversion speeds. In this sense, the three event dates also help to illustrate the regime interpretation. The COVID date still lies in the pre-2022 cloud and therefore mainly serves as a reference point for the earlier calibration geometry. The ECB hike date falls into the transition period, whereas the ECB cut date belongs to the later post-2022 cluster in which local instability can persist despite low pricing errors.
	
	\item \textbf{Parameter uncertainty and risk.}
	The fan chart for the second mean-reversion speed \(a_y\) in Figure~\ref{fig:ci_fan_ay} adds a final layer to this picture. For long stretches, \(\hat a_y\) lies close to the lower part of the admissible region and the confidence intervals are asymmetric. This confirms that naive Gaussian intervals on the original scale are inappropriate near parameter boundaries. The robust MAD-based scaling used in the main text avoids the inflated full-range bands that arise under classical MSE-based scaling.
	
	At the same time, episodes in which the interval collapses to the point estimate must not be misread as unusually high precision. Under rank deficiency, the Moore--Penrose pseudo-inverse removes non-estimable directions from the covariance approximation rather than inflating them. Parameter uncertainty is highly time-varying and boundary-sensitive. It is closely linked to local identifiability. Treating the daily calibration \(\hat\Pi_t\) as a precise input can be misleading even when the RMSRE is low. The confidence bands should therefore be read jointly with the leverage, influence, and Effective Degrees of Freedom diagnostics. For the confidence intervals, the event dates mainly serve as common reference points rather than as the dates with the largest uncertainty bands.
	
\end{description}

\paragraph{Limitations and Future Research}
While the framework is theoretically applicable to general least-squares calibration problems, the present implementation relies on the availability of closed-form gradients for interest rate caps. An important challenge remains for instruments such as swaptions in the G2++ model or path-dependent derivatives, where prices require numerical integration or simulation. In these cases, the Jacobian must be approximated by finite differences, algorithmic differentiation, or other numerical differentiation schemes. Future research should therefore focus on derivative approximations that are sufficiently stable to support the linear systems involving the Fisher Information Matrix.

These diagnostics can also be used to improve calibration rather than merely describe it ex-post. For instance, leverage scores could be used to construct leverage-balanced weighting schemes, where weights are adjusted to prevent individual maturities from dominating the local fit. Alternatively, in actuarial applications such as long-term pension projections, the weighting scheme could be designed to emphasize maturities that are particularly relevant for the liability profile, such as the $20$Y--$30$Y segment. More generally, days flagged by boundary leverage, influence spikes, or rank loss could be subjected to additional stress checks, for example by rerunning the calibration after removing or reweighting the dominant maturity.

In an actuarial workflow, these diagnostics should be understood as one layer of model governance, not as a complete validation procedure. In this sense, the diagnostics are descriptive and can also provide practical rules for deciding which calibrations require additional scrutiny.

A further open question concerns the theoretical status of the empirically observed boundary-dominated leverage profile. In this paper, boundary dominance is documented empirically and interpreted as geometrically plausible in the G2++ cap calibration problem. A full mathematical characterization of when and why leverage tends to concentrate at short- and long-end maturities, however, goes beyond the scope of this paper. Developing such a theory would be a relevant direction for future research, as it would connect the empirical diagnostics more directly to the local sensitivity geometry of the calibration problem.

\FloatBarrier
\section*{Data and Code Availability}
The market data used in this study were obtained from LSEG Workspace and associated OIS curve sources and are subject to third-party licensing restrictions. Derived project data supporting the figures and diagnostics, as well as the R code used for the calibration diagnostics and figure generation, are available from the corresponding author on reasonable request.

\FloatBarrier
\section*{Author Contributions}

Philipp Mahler developed the methodology, implemented the empirical analysis, and prepared the manuscript. Peter Ruckdeschel contributed to the conceptual design, interpretation of the statistical calibration framework, and revision of the manuscript. Both authors read and approved the final manuscript.

\FloatBarrier
\clearpage

\FloatBarrier
\begin{appendices}

\section{Appendix}

\subsection{Detailed Mathematical Derivations}\label{app:explicit_derivatives}
This section provides the detailed step-by-step derivations for the key components used in the main body of the paper, including the first and second-order derivatives of the G2++ cap pricing formula.

\subsubsection{Derivation of the Cap Price First-Order Derivatives}\label{app:first_order_derivatives}
The goal is to compute the Jacobian matrix \( \mathbf{J} \), whose entries are the first-order partial derivatives of each cap price with respect to each model parameter \( p \in \Pi \). The derivative of a cap is the sum of the derivatives of its constituent caplets.

We proceed by explicitly calculating the terms of the chain rule expansion introduced in Equation~\eqref{eq:chain_rule_expansion}.

\paragraph{Step 1: The Outer Derivative \texorpdfstring{$\partial \text{Caplet} / \partial \Phi$}{dCaplet/dPhi}}
Since the notional and pricing factors are constants with respect to \( p \), the differential operator acts only on the \( \Phi \) terms of the caplet pricing formula:
\begin{equation*}
	\frac{\partial \text{Caplet}_i}{\partial p} = N \left[ P(0,t_{i-1})\frac{\partial \Phi(h_{+,i})}{\partial p} - \bigl(1+K\delta_i\bigr)P(0,t_i)\frac{\partial \Phi(h_{-,i})}{\partial p} \right].
\end{equation*}

\paragraph{Step 2: Derivative of the Normal CDF \texorpdfstring{$\partial \Phi / \partial h$}{dPhi/dh}}
Using the standard property that $\frac{\partial \Phi(x)}{\partial x} = \varphi(x)$ (the standard normal PDF), we apply the chain rule to the inner terms:
\begin{equation*}
	\frac{\partial \Phi(h_{\pm,i})}{\partial p} = \varphi(h_{\pm,i}) \cdot \frac{\partial h_{\pm,i}}{\partial p}.
\end{equation*}

\paragraph{Step 3: Substitution of the Derivative of \texorpdfstring{$h_{\pm}$}{h}}
Next, we substitute the derivative of \( h_{\pm,i} \) with respect to \(p\), which is given by
\begin{equation*}
	\frac{\partial h_{\pm,i}}{\partial p} = \left( -\frac{L_i}{\Sigma_i^2} \pm \frac{1}{2} \right)\frac{\partial \Sigma_i}{\partial p},
\end{equation*}
where
\[
L_i = \log\left(\frac{P(0,t_{i-1})}{(1+K\delta_i)P(0,t_i)}\right).
\]
This introduces \( \frac{\partial \Sigma_i}{\partial p} \) as a common factor.

\paragraph{Step 4: Transition from \texorpdfstring{$\Sigma$}{Sigma} to \texorpdfstring{$\Sigma^2$}{Sigma-squared}}
The final algebraic step is to transition from the derivative of \( \Sigma_i \) to the derivative of \( \Sigma_i^2 \), as the latter is algebraically simpler to calculate. We use the relationship
\begin{equation*}
	\frac{\partial \Sigma_i}{\partial p} = \frac{1}{2\Sigma_i}\frac{\partial \Sigma_i^2}{\partial p}.
\end{equation*}
Substituting this into the expression from Step 3 yields a formula in which \( \frac{\partial \Sigma_i^2}{\partial p} \) appears as a common factor.

\paragraph{Step 5: Factorization and Definition of \texorpdfstring{$\mathcal{M}_i(\Pi)$}{Mi(Pi)}}
By factoring out the common term \( \frac{\partial \Sigma_i^2}{\partial p} \), we obtain
\begin{align*}
	\mathcal{M}_i(\Pi)
	=
	\frac{N}{2\Sigma_i} \bigg[
	& P(0,t_{i-1})\varphi(h_{+,i}) \left(-\frac{L_i}{\Sigma_i^2} + \frac{1}{2}\right) \\
	& - \bigl(1+K\delta_i\bigr)P(0,t_i)\varphi(h_{-,i}) \left(-\frac{L_i}{\Sigma_i^2} - \frac{1}{2}\right)
	\bigg],
\end{align*}
and therefore
\begin{equation*}
	\frac{\partial \text{Caplet}_i}{\partial p}
	=
	\mathcal{M}_i(\Pi)\cdot \frac{\partial \Sigma_i^2}{\partial p}.
\end{equation*}
This factorization is the main computational device for an efficient implementation, separating the strike-dependent factor \( \mathcal{M}_i(\Pi) \) from the reusable variance-to-parameter sensitivity \( \frac{\partial \Sigma_i^2}{\partial p} \).

\paragraph{Step 6: Exact Simplification of \texorpdfstring{$\mathcal{M}_i(\Pi)$}{Mi(Pi)}}
The expanded representation above simplifies exactly. Since
\[
h_{+,i}-h_{-,i}=\Sigma_i,
\qquad
h_{+,i}+h_{-,i}=\frac{2L_i}{\Sigma_i},
\]
we obtain
\[
h_{+,i}^2-h_{-,i}^2 = 2L_i.
\]
Hence
\[
\frac{\varphi(h_{+,i})}{\varphi(h_{-,i})}
=
\exp\!\left(-\frac{h_{+,i}^2-h_{-,i}^2}{2}\right)
=
e^{-L_i}.
\]
Using the definition of \(L_i\), this implies
\[
P(0,t_{i-1})\varphi(h_{+,i})
=
\bigl(1+K\delta_i\bigr)P(0,t_i)\varphi(h_{-,i}).
\]
Substituting this identity into the expanded expression for \( \mathcal{M}_i(\Pi) \) yields the exact short form stated in Equation~\eqref{eq:factor_M}, together with the equivalent representation
\begin{equation*}
	\mathcal{M}_i(\Pi)
	=
	\frac{N\,P(0,t_{i-1})\varphi(h_{+,i})}{2\Sigma_i}
	=
	\frac{N\bigl(1+K\delta_i\bigr)P(0,t_i)\varphi(h_{-,i})}{2\Sigma_i}.
\end{equation*}

\subsubsection{Final Expressions for Parameter Sensitivities}\label{app:final_expressions}
The derivative of the total cap price is the sum of the derivatives of its constituent caplets. By combining the \( \mathcal{M}_i \) factor with the analytical derivatives of the variance function \( \Sigma^2 \), we obtain the final expressions for each entry in the Jacobian. For brevity, we denote \( C(\alpha, t_{i-1}, t_i) \) by \( C_i(\alpha) \). Its derivative with respect to the first argument is denoted by \( C_i'(\alpha) \), where
\[
C_i'(\alpha) = \frac{\partial C(\alpha, t_{i-1}, t_i)}{\partial \alpha}
= \frac{e^{-\alpha(t_i - t_{i-1})} \left( \alpha(t_i - t_{i-1}) + 1 \right) - 1}{\alpha^2}.
\]
\paragraph{Derivative with respect to \( \sigma_x \):}
\begin{align*}
	\frac{\partial \text{Cap}}{\partial \sigma_x} = \sum_{i} \mathcal{M}_i(\Pi) \cdot \bigg[ & 2\sigma_x C_i(a_x)^2 C(2a_x, 0, t_{i-1}) \\
	& + 2\rho_{xy}\sigma_y C_i(a_x)C_i(a_y)C(a_x+a_y, 0, t_{i-1}) \bigg].
\end{align*}

\paragraph{Derivative with respect to \( \sigma_y \):}
\begin{align*}
	\frac{\partial \text{Cap}}{\partial \sigma_y} = \sum_{i} \mathcal{M}_i(\Pi) \cdot \bigg[ & 2\sigma_y C_i(a_y)^2 C(2a_y, 0, t_{i-1}) \\
	& + 2\rho_{xy}\sigma_x C_i(a_x)C_i(a_y)C(a_x+a_y, 0, t_{i-1}) \bigg]
\end{align*}

\paragraph{Derivative with respect to \( \rho_{xy} \):}
\[
\frac{\partial \text{Cap}}{\partial \rho_{xy}} = \sum_{i} \mathcal{M}_i(\Pi) \cdot \left[ 2\sigma_x\sigma_y C_i(a_x)C_i(a_y)C(a_x+a_y, 0, t_{i-1}) \right]
\]

\paragraph{Derivative with respect to \( a_x \):}
\begin{align*}
	\frac{\partial \text{Cap}}{\partial a_x} = \sum_{i} \mathcal{M}_i(\Pi) \cdot \bigg[ & \sigma_x^2 ( 2C_i(a_x)C'_i(a_x)C(2a_x, 0, t_{i-1}) \\
	& + C_i(a_x)^2 C'(2a_x, 0, t_{i-1}) \cdot 2 ) \\
	& + 2\rho_{xy}\sigma_x\sigma_y ( C'_i(a_x)C_i(a_y)C(a_x+a_y, 0, t_{i-1}) \\ 
	& + C_i(a_x)C_i(a_y)C'(a_x+a_y, 0, t_{i-1}) ) \bigg]
\end{align*}

\paragraph{Derivative with respect to \( a_y \):}
\begin{align*}
	\frac{\partial \text{Cap}}{\partial a_y} = \sum_{i} \mathcal{M}_i(\Pi) \cdot \bigg[ & \sigma_y^2 ( 2C_i(a_y)C'_i(a_y)C(2a_y, 0, t_{i-1}) \\ 
	& + C_i(a_y)^2 C'(2a_y, 0, t_{i-1}) \cdot 2 ) \\
	& + 2\rho_{xy}\sigma_x\sigma_y ( C_i(a_x)C'_i(a_y)C(a_x+a_y, 0, t_{i-1}) \\ 
	& + C_i(a_x)C_i(a_y)C'(a_x+a_y, 0, t_{i-1}) ) \bigg]
\end{align*}

\subsubsection{No-Arbitrage Derivation of the Cap-Floor Parity}
\label{app:cap_floor_parity}

In Section~\ref{subsec:jacobian_derivation}, we rely on the exact geometric equivalence between caps, floors, and payer swaps to establish the universality of our diagnostic framework. Here, we provide the step-by-step no-arbitrage derivation of this parity.

At its core, the put--call parity relies on a simple algebraic identity for the maximum function. For any two real numbers $L$ and $K$, the difference between the positive part of $(L-K)$ and $(K-L)$ is simply the linear difference:
\begin{equation}
	\label{eq:fundamental_identity}
	\max(L - K, 0) - \max(K - L, 0) = L - K.
\end{equation}

Analogously to Eq.~\eqref{eq:caplet_payoff}, the payoff of a \textit{floorlet} at time $t_i$ is:
\begin{equation}
	\text{Payoff}_\text{Floorlet}(t_i) = N \delta_i \max(K - L(t_{i-1}, t_i), 0).
\end{equation}
Using the identity~\eqref{eq:fundamental_identity}, the difference between the payoffs at time $t_i$ is strictly linear:
\begin{equation}
	\label{eq:fra_payoff}
	\text{Caplet}_i(t_i) - \text{Floorlet}_i(t_i) = N \delta_i \big( L(t_{i-1}, t_i) - K \big).
\end{equation}

To find the present value at $t=0$, we take the expectation under the $t_i$-forward measure $\mathbb{Q}^{t_i}$ with numeraire $P(\cdot, t_i)$. Since the forward rate $L(\cdot, t_{i-1}, t_i)$ is a martingale under $\mathbb{Q}^{t_i}$, the expectation simplifies to today's forward rate $L(0, t_{i-1}, t_i)$:
\begin{equation}
	\text{Caplet}_i(0) - \text{Floorlet}_i(0) = P(0, t_i) N \delta_i \big( L(0, t_{i-1}, t_i) - K \big).
\end{equation}
We eliminate the forward rate by substituting the standard no-arbitrage relation $L(0, t_{i-1}, t_i) = \frac{1}{\delta_i} \left( \frac{P(0, t_{i-1})}{P(0, t_i)} - 1 \right)$, which yields:
\begin{equation}
	\label{eq:caplet_floorlet_parity}
	\text{Caplet}_i(0) - \text{Floorlet}_i(0) = N \Big[ P(0, t_{i-1}) - (1 + K \delta_i)P(0, t_i) \Big]. 
\end{equation}
Summing Eq.~\eqref{eq:caplet_floorlet_parity} over all payment periods $i = m+1, \dots, n$ yields the difference between a cap and a floor. This aggregate difference matches precisely the present value of a payer swap receiving the floating rate and paying the fixed rate $K$:
\begin{equation}
	\text{Cap}(0, \mathcal{T}, N, K) - \text{Floor}(0, \mathcal{T}, N, K) = \text{PayerSwap}(0, \mathcal{T}, N, K).
\end{equation}
By separating the summation into a floating and a fixed leg, the floating leg forms a telescoping sum, providing the explicit, model-independent swap value:
\begin{align}
	\text{PayerSwap}(0, \mathcal{T}, N, K) &= \sum_{i=m+1}^n N \Big[ P(0, t_{i-1}) - P(0, t_i) \Big] - \sum_{i=m+1}^n N K \delta_i P(0, t_i) \nonumber \\
	&= N \Big[ P(0, t_m) - P(0, t_n) \Big] - N \cdot K \sum_{i=m+1}^n \delta_i P(0, t_i). \label{eq:swap_final_app}
\end{align}

\subsubsection{Analytical Derivation of the Floor Jacobian}
\label{app:floor_jacobian}

To rigorously confirm the geometric equivalence of caps and floors stated in the main text, we explicitly derive the Jacobian matrix for an interest rate floor and demonstrate its algebraic identity to the cap Jacobian.

In the G2++ model, a floorlet for the period $[t_{i-1}, t_i]$ is mathematically equivalent to a European Call Option on the zero-coupon bond $P(t_{i-1}, t_i)$. The pricing formula for a floorlet at $t=0$ is given by:
\begin{equation}
	\text{Floorlet}_i = N \Big[ (1 + K\delta_i)P(0, t_i)\Phi(-h_{-,i}) - P(0, t_{i-1})\Phi(-h_{+,i}) \Big],
\end{equation}
where the functions $h_{+,i}$ and $h_{-,i}$ are defined exactly as in the caplet pricing formula (see Eq.~\ref{eq:h_pm}).

We compute the partial derivative of the floorlet price with respect to an arbitrary model parameter $p \in \Pi$. Since the notional $N$, the strike $K$, the accrual period $\delta_i$, and the initial discount factors $P(0, t)$ are deterministically fixed and independent of the stochastic parameters $p$, the differential operator solely acts on the cumulative standard normal distribution functions $\Phi$. Applying the chain rule yields:
\begin{equation}
	\frac{\partial \text{Floorlet}_i}{\partial p} = N \left[ (1 + K\delta_i)P(0, t_i) \varphi(-h_{-,i}) \left(-\frac{\partial h_{-,i}}{\partial p}\right) - P(0, t_{i-1}) \varphi(-h_{+,i}) \left(-\frac{\partial h_{+,i}}{\partial p}\right) \right],
\end{equation}
where $\varphi(\cdot)$ denotes the standard normal probability density function.

Exploiting the symmetry property of the standard normal density, $\varphi(-x) = \varphi(x)$, we can factor out the negative signs and rearrange the terms:
\begin{align}
	\frac{\partial \text{Floorlet}_i}{\partial p} &= N \left[ -(1 + K\delta_i)P(0, t_i) \varphi(h_{-,i}) \frac{\partial h_{-,i}}{\partial p} + P(0, t_{i-1}) \varphi(h_{+,i}) \frac{\partial h_{+,i}}{\partial p} \right] \nonumber \\
	&= N \left[ P(0, t_{i-1}) \varphi(h_{+,i}) \frac{\partial h_{+,i}}{\partial p} - (1 + K\delta_i)P(0, t_i) \varphi(h_{-,i}) \frac{\partial h_{-,i}}{\partial p} \right].
\end{align}
Comparing this result to the derivation in Appendix~\ref{app:first_order_derivatives}, we see that this final expression is algebraically identical to the derivative of the corresponding caplet:
\begin{equation}
	\frac{\partial \text{Floorlet}_i}{\partial p} \equiv \frac{\partial \text{Caplet}_i}{\partial p}.
\end{equation}

Consequently, the price-to-variance sensitivity factor for the floorlet precisely matches that of the caplet:
\begin{equation}
	\mathcal{M}_i^{\text{Floor}}(\Pi) = \mathcal{M}_i^{\text{Cap}}(\Pi).
\end{equation}
Summing these derivatives over all caplets and floorlets within the respective payment schedules directly implies $\mathbf{J}_{\text{Floor}} = \mathbf{J}_{\text{Cap}}$. This proves that the tangent space and the structural geometric properties of the calibration problem are perfectly invariant between caps and floors.

\subsubsection{Derivation of Second-Order Derivatives (Hessian)}\label{app:second_order_derivatives}
For second-order optimization algorithms (such as Newton--Raphson) and for advanced local diagnostics, the Hessian matrix of the cap price is required. At the caplet level, the entry
\[
\frac{\partial^2 \text{Caplet}_i}{\partial p_u \partial p_v}
\]
is obtained by differentiating the factorized first-order representation.

\paragraph{General Approach via the Product Rule}
Since
\[
\frac{\partial \text{Caplet}_i}{\partial p_v}
=
\mathcal{M}_i(\Pi)\frac{\partial \Sigma_i^2}{\partial p_v},
\]
the product rule yields
\begin{equation*}
	\frac{\partial^2 \text{Caplet}_i}{\partial p_u \partial p_v}
	=
	\frac{\partial \mathcal{M}_i(\Pi)}{\partial p_u}\frac{\partial \Sigma_i^2}{\partial p_v}
	+
	\mathcal{M}_i(\Pi)\frac{\partial^2 \Sigma_i^2}{\partial p_u \partial p_v}.
\end{equation*}
Since $\mathcal{M}_i(\Pi)$ depends on the model parameters only through $\Sigma_i$, we may write
\[
\frac{\partial \mathcal{M}_i(\Pi)}{\partial p_u}
=
\mathcal{M}_i'(\Pi)\frac{\partial \Sigma_i^2}{\partial p_u},
\qquad
\mathcal{M}_i'(\Pi):=\frac{1}{2\Sigma_i}\frac{d\mathcal{M}_i}{d\Sigma_i}.
\]
Hence
\begin{equation*}
	\frac{\partial^2 \text{Caplet}_i}{\partial p_u \partial p_v}
	=
	\mathcal{M}_i'(\Pi)
	\frac{\partial \Sigma_i^2}{\partial p_u}
	\frac{\partial \Sigma_i^2}{\partial p_v}
	+
	\mathcal{M}_i(\Pi)\frac{\partial^2 \Sigma_i^2}{\partial p_u \partial p_v}.
\end{equation*}

\paragraph{Derivative of the Weighting Factor}
Starting from the exact short form in Eq.~\eqref{eq:factor_M},
\[
\mathcal{M}_i(\Pi)
=
\frac{N\,P(0,t_{i-1})\varphi(h_{+,i})}{2\Sigma_i},
\]
a direct differentiation with respect to \(\Sigma_i\) yields a compact expression. The key identities are
\[
\varphi'(x)=-x\varphi(x)
\qquad\text{and}\qquad
\frac{d h_{+,i}}{d\Sigma_i}
=
-\frac{L_i}{\Sigma_i^2}+\frac12
=
-\frac{h_{-,i}}{\Sigma_i}.
\]
Using these relations, one obtains
\begin{equation*}
	\mathcal{M}_i'(\Pi)
	=
	\frac{N\,P(0,t_{i-1})\varphi(h_{+,i})}{4\Sigma_i^3}
	\bigl(h_{+,i}h_{-,i}-1\bigr).
\end{equation*}

\paragraph{Second Partial Derivatives of the Variance Function}
The remaining component is the Hessian of the variance function $\Sigma_i^2$. Since $\Sigma_i^2$ is smooth, Schwarz's theorem applies, so
\[
\frac{\partial^2 \Sigma_i^2}{\partial p_u \partial p_v}
=
\frac{\partial^2 \Sigma_i^2}{\partial p_v \partial p_u}.
\]
The resulting expressions involve first and second derivatives of the auxiliary function \(C(\alpha,t,T)\). Since the Hessian is symmetric, it is sufficient to derive the upper triangular entries only. We categorize these derivations based on the type of parameters involved.

\subsubsection*{Auxiliary Functions}
For better readability, we recall the auxiliary function
\[
C(\alpha,t,T)=\frac{1-e^{-\alpha(T-t)}}{\alpha}
\]
and define its derivatives with respect to \(\alpha\). With the previously defined accrual period \(\delta_i=t_i-t_{i-1}\), we have
\begin{align*}
	C_i(\alpha) &= C(\alpha,t_{i-1},t_i),\\
	C_i'(\alpha) &= \frac{\partial C_i(\alpha)}{\partial \alpha}
	= \frac{e^{-\alpha\delta_i}(\alpha\delta_i+1)-1}{\alpha^2},\\
	C_i''(\alpha) &= \frac{\partial^2 C_i(\alpha)}{\partial \alpha^2}
	= \frac{2-e^{-\alpha\delta_i}\bigl((\alpha\delta_i+1)^2+1\bigr)}{\alpha^3}.
\end{align*}
Additionally, we define
\begin{align*}
	\Gamma_x &= C(2a_x,0,t_{i-1}),\\
	\Gamma_y &= C(2a_y,0,t_{i-1}),\\
	\Gamma_{xy} &= C(a_x+a_y,0,t_{i-1}).
\end{align*}
These quantities also depend on the mean-reversion parameters and must therefore be differentiated analogously. For example,
\[
\frac{\partial \Gamma_x}{\partial a_x}
=
2\,C'(2a_x,0,t_{i-1}).
\]

\subsubsection*{Derivatives with respect to Volatility and Correlation}
These derivatives are the simplest, as the variance function $\Sigma^2$ is quadratic in volatilities and linear in correlation.

\textbf{Second derivatives of volatilities:}
\begin{align*}
	\frac{\partial^2 \Sigma_i^2}{\partial \sigma_x^2} &= 2 C_i(a_x)^2 \Gamma_x ,\\
	\frac{\partial^2 \Sigma_i^2}{\partial \sigma_y^2} &= 2 C_i(a_y)^2 \Gamma_y .
\end{align*}

\textbf{Mixed volatility and correlation derivatives:}
\begin{align*}
	\frac{\partial^2 \Sigma_i^2}{\partial \sigma_x \partial \sigma_y} &= 2 \rho_{xy} C_i(a_x) C_i(a_y) \Gamma_{xy}, \\
	\frac{\partial^2 \Sigma_i^2}{\partial \sigma_x \partial \rho_{xy}} &= 2 \sigma_y C_i(a_x) C_i(a_y) \Gamma_{xy}, \\
	\frac{\partial^2 \Sigma_i^2}{\partial \sigma_y \partial \rho_{xy}} &= 2 \sigma_x C_i(a_x) C_i(a_y) \Gamma_{xy}.
\end{align*}

\textbf{Second derivative of correlation:}
\[
\frac{\partial^2 \Sigma_i^2}{\partial \rho_{xy}^2} = 0.
\]

\subsubsection*{Derivatives containing Mean Reversion Speeds}
Now, we turn to derivatives involving the mean reversion speeds \(a_x\) and \(a_y\). We first list the mixed derivatives involving one mean reversion parameter. Several additional expressions follow by symmetry under the exchange of \(x\) and \(y\), but the cross-interactions are stated explicitly where needed.
\begin{align*}
	\frac{\partial^2 \Sigma_i^2}{\partial \sigma_x \partial a_x} &= 2\sigma_x \left[ 2 C_i(a_x) C'_i(a_x) \Gamma_x + C_i(a_x)^2 \frac{\partial \Gamma_x}{\partial a_x} \right] \\
	& \quad + 2\rho_{xy}\sigma_y \left[ C'_i(a_x) C_i(a_y) \Gamma_{xy} + C_i(a_x) C_i(a_y) \frac{\partial \Gamma_{xy}}{\partial a_x} \right], \\[1em]
	\frac{\partial^2 \Sigma_i^2}{\partial \sigma_y \partial a_x}
	&=
	2\rho_{xy}\sigma_x
	\left[
	C_i'(a_x)C_i(a_y)\Gamma_{xy}
	+
	C_i(a_x)C_i(a_y)\frac{\partial \Gamma_{xy}}{\partial a_x}
	\right],\\
	\frac{\partial^2 \Sigma_i^2}{\partial \sigma_x \partial a_y}
	&=
	2\rho_{xy}\sigma_y
	\left[
	C_i(a_x)C_i'(a_y)\Gamma_{xy}
	+
	C_i(a_x)C_i(a_y)\frac{\partial \Gamma_{xy}}{\partial a_y}
	\right],\\
	\frac{\partial^2 \Sigma_i^2}{\partial \rho_{xy} \partial a_x} &= 2\sigma_x \sigma_y \left[ C'_i(a_x) C_i(a_y) \Gamma_{xy} + C_i(a_x) C_i(a_y) \frac{\partial \Gamma_{xy}}{\partial a_x} \right].
\end{align*}

The most involved case is the second derivative with respect to the mean-reversion speeds.

\textbf{Second derivative with respect to \( a_x \):}
\begin{align*}
	\frac{\partial^2 \Sigma_i^2}{\partial a_x^2} &= \sigma_x^2 \Bigg[ 2\left( (C'_i(a_x))^2 + C_i(a_x)C''_i(a_x) \right) \Gamma_x \\
	&\phantom{{}= \sigma_x^2 \Bigg[} + 4 C_i(a_x) C'_i(a_x) \frac{\partial \Gamma_x}{\partial a_x} + C_i(a_x)^2 \frac{\partial^2 \Gamma_x}{\partial a_x^2} \Bigg] \\
	&\quad + 2\rho_{xy}\sigma_x\sigma_y \Bigg[ C''_i(a_x) C_i(a_y) \Gamma_{xy} + 2 C'_i(a_x) C_i(a_y) \frac{\partial \Gamma_{xy}}{\partial a_x} \\
	&\phantom{{}\quad + 2\rho_{xy}\sigma_x\sigma_y \Bigg[} + C_i(a_x) C_i(a_y) \frac{\partial^2 \Gamma_{xy}}{\partial a_x^2} \Bigg].
\end{align*}

\textbf{Mixed derivative \( a_x, a_y \):}
Since the $x$-term of $\Sigma^2$ depends only on $a_x$ and the $y$-term only on $a_y$, the only interaction occurs in the correlation term.
\begin{align*}
	\frac{\partial^2 \Sigma_i^2}{\partial a_x \partial a_y} &= 2\rho_{xy}\sigma_x\sigma_y \Bigg[ C'_i(a_x) C'_i(a_y) \Gamma_{xy} + C'_i(a_x) C_i(a_y) \frac{\partial \Gamma_{xy}}{\partial a_y} \\
	& \qquad \qquad \quad + C_i(a_x) C'_i(a_y) \frac{\partial \Gamma_{xy}}{\partial a_x} + C_i(a_x) C_i(a_y) \frac{\partial^2 \Gamma_{xy}}{\partial a_x \partial a_y} \Bigg].
\end{align*}

Note that for the term $\Gamma_{xy} = C(a_x+a_y, 0, t_{i-1})$, the partial derivatives are straightforward:
\[
\frac{\partial \Gamma_{xy}}{\partial a_x} = \frac{\partial \Gamma_{xy}}{\partial a_y} = C'(a_x+a_y, 0, t_{i-1}).
\]

\subsubsection{Empirical adequacy of the Gauss--Newton approximation}
\label{subsec:gn_adequacy}

Since our diagnostics are based on the local Jacobian geometry, it is natural to ask
whether a first-order approximation is sufficiently accurate for the G2++ cap calibration problem.
At the calibrated parameter \(\hat{\Pi}\), the exact Hessian of the weighted least-squares objective
\[
Q(\Pi)=\sum_{k=1}^m w_k\bigl(y_k-g_k(\Pi)\bigr)^2
\]
can be written as
\[
\nabla^2 Q(\hat{\Pi})
=
2\mathbf{J}(\hat{\Pi})^\top \mathbf{W}\mathbf{J}(\hat{\Pi})
+
\Delta_{\mathrm{GN}}(\hat{\Pi}),
\]
where
\[
\Delta_{\mathrm{GN}}(\hat{\Pi})
:=
-\,2\sum_{k=1}^m w_k\,r_k(\hat{\Pi})\,\nabla^2 g_k(\hat{\Pi}),
\qquad
r_k(\hat{\Pi})=y_k-g_k(\hat{\Pi}).
\]
Hence, \(\Delta_{\mathrm{GN}}(\hat{\Pi})\) is the residual-weighted second-order correction term measuring the deviation of the exact objective Hessian from its Gauss--Newton approximation.

We evaluated this correction term numerically over the full sample. Although the individual
cap price functions are clearly nonlinear, \(\Delta_{\mathrm{GN}}(\hat{\Pi})\) is norm-wise negligible relative
to the Gauss--Newton term at the calibrated solutions. In Frobenius norm, the ratio
\[
\frac{\|\Delta_{\mathrm{GN}}(\hat{\Pi})\|_F}
{\|2\mathbf{J}(\hat{\Pi})^\top \mathbf{W}\mathbf{J}(\hat{\Pi})\|_F}
\]
has mean \(0.19\%\), median \(0.012\%\), and 90\% quantile \(0.63\%\). Therefore, for the local geometry relevant to our diagnostics, \(\mathbf{J}^\top \mathbf{W}\mathbf{J}\) provides an essentially complete curvature description. In this sense, the linearized tangent-space approximation describes the local calibration geometry very well in our sample. Correspondingly, near the calibrated solutions, Newton--Raphson and Gauss--Newton updates are empirically nearly indistinguishable, since the residual-weighted second-order correction is negligible.

\subsection{Theoretical Foundations: Influence Functions and Asymptotic Normality}
\label{app:theoretical_foundations}

As discussed in Section~\ref{subsec:literature_review}, we interpret the calibration estimator within the framework of robust statistics \cite{huber1981robust, hampel1986robust}. In this section, we provide the formal definitions for the Influence Function and the resulting asymptotic normality, adopting the notation and setup of Rieder \cite{rieder1994robust} and van der Vaart \cite{van1998asymptotic}.

Consistent with this framework, the estimator $\hat{\Pi}$ is asymptotically normal. Under this setup, the covariance matrix of the estimator is given by the inverse of the Fisher Information Matrix (FIM):
\begin{equation}
	\widehat{\text{Cov}}(\hat{\Pi}) \approx \mathcal{I}(\hat{\Pi})^{-1} = \sigma^2 (\mathbf{J}^\top \mathbf{W} \mathbf{J})^{-1},
\end{equation}
where \( \sigma^2 \) is the estimated variance of the residuals.

\paragraph{Estimators as Statistical Functionals}
In robust statistics, we view an estimator not merely as a function of the sample data, but as a functional $T$ defined on a space of probability measures $\mathcal{M}$. Let $F_{\Pi}$ be the model distribution parameterized by $\Pi \in \Theta \subset \mathbb{R}^p$, and let $G_n$ be the empirical distribution function of the observed data. An estimator $\hat{\Pi}_n$ is defined as the value of the functional at the empirical distribution:
\begin{equation}
	\hat{\Pi}_n = T(G_n).
\end{equation}
Fisher consistency requires that at the ideal model distribution $F_{\Pi}$, the functional recovers the true parameter: $T(F_{\Pi}) = \Pi$.

\paragraph{The Influence Function via Gâteaux Differentiation}
Because our calibration minimizes the WLS objective, \(\hat{\Pi}_n\) can be viewed as an \(M\)-estimator. At the same time, the first-order conditions for this minimization problem imply that \(\hat{\Pi}_n\) solves a system of estimating equations, formally classifying it also as a \(Z\)-estimator. To analyze the stability of this estimator \(T\) under data contamination, we use the functional derivative viewpoint standard in robust asymptotic statistics. The Influence Function is defined as the Gâteaux derivative of \(T\) at \(F\) in the direction of a Dirac measure \(\delta_x\):
\begin{equation}
	\text{IF}(x; T, F) := \lim_{\varepsilon \downarrow 0} \frac{ T((1-\varepsilon)F + \varepsilon \delta_x) - T(F)}{\varepsilon}.
\end{equation}
This derivative quantifies the infinitesimal effect of a single observation \(x\) on the estimator. According to van der Vaart \cite[Ch.~20]{van1998asymptotic}, under Hadamard differentiability of the functional \(T\) and the usual regularity conditions ensuring asymptotic linearity of the corresponding \(Z\)-estimator, this yields the standard asymptotic linear representation
\begin{equation}
	\label{ALErep}
	\sqrt{n}(\hat{\Pi}_n - \Pi) = \frac{1}{\sqrt{n}} \sum_{i=1}^n\text{IF}(x_i; T, F) + o_p(1).
\end{equation}

For a general \(M\)- or \(Z\)-estimator, the relevant object is the estimating function \(\varphi_\theta=\partial \rho_\theta/\partial\theta\), where \(\rho_\theta\) coincides with the negative, possibly weighted, log-likelihood only in the maximum-likelihood case. Hence, \(\varphi_\theta\) coincides with the likelihood score \(\Lambda_\theta=\partial\log p_\theta/\partial\theta\), up to sign convention, only in this special case. In general, the normalization by the inverse Fisher Information Matrix is replaced by the inverse sensitivity matrix
\[
D_\theta = \left(\mathbb{E}_\theta[\varphi_\theta \Lambda_\theta^\top]\right)^{-1}.
\] 
The vector formulation derived in Section~\ref{subsec:diagnostic_tools} is the finite-sample empirical realization of this theoretical concept.

\paragraph{Asymptotic Normality}
The asymptotic linear representation in Eq.~\eqref{ALErep} implies that the estimator is asymptotically normal:
\begin{equation}
	\sqrt{n}(\hat{\Pi}_n - \Pi_0) \xrightarrow{d} \mathcal{N}\left(0, \mathcal{I}(\Pi_0)^{-1}\right).
\end{equation}
Thus, the local covariance structure is governed by the inverse Fisher Information Matrix in this local WLS sense. Here, this matrix is approximated by \(\mathbf{J}^\top\mathbf{W}\mathbf{J}\) conditional on the realized weights and calibration design, which is why it enters both the uncertainty quantification and the influence-based diagnostics. Near active constraints or rank-deficient cases, the resulting intervals should be read as local diagnostic summaries rather than exact probabilistic guarantees.

\paragraph{Remark on One-Step Corrections}
A useful asymptotic interpretation is provided by the one-step theorem; see van der Vaart \cite[Sec.~5.7, Thm.~5.45]{van1998asymptotic}. In smooth finite-dimensional estimation problems, a single Newton- or scoring-type correction applied to a sufficiently accurate preliminary estimator is already asymptotically equivalent to the fully iterated solution. In particular, once the starting estimator is consistent with a suitable rate, the one-step correction inherits the first-order asymptotic distribution of the target estimator; to first order, the asymptotic covariance is then governed by the local derivative at the target rather than by the preliminary covariance. In our setting, the empirical near-equivalence between the exact Hessian and its Gauss--Newton approximation makes this one-step intuition locally plausible, although we do not use it as a formal basis for the diagnostics.

\paragraph{Conceptual Relation to xAI}
There are loose conceptual parallels to model-interpretability methods in machine learning; see Molnar \cite{molnar2022interpretable} for an overview. Our local linearization, leverage, and influence diagnostics aim to make a complex input--output relation locally interpretable. The analogy is limited, however: unlike the predominantly model-agnostic xAI literature, our framework is analytic, parameter-structural, and tailored to the constrained geometry of weighted calibration rather than to generic black-box explanation. We therefore do not pursue SHAP-type additive decompositions here, since our goal is not a generic black-box explanation, but local parameter-geometric sensitivity.

\subsection{The Geometric Interpretation of the Hat Matrix}
\label{app:geometric_interpretation}

The statistical diagnostics used in this paper have a geometric interpretation that clarifies the role of each market observation. The core task of calibration is to find the model parameters \( \hat{\Pi} \) such that the vector of model prices \( \mathbf{g}(\hat{\Pi}) \) best matches the observed market prices \( \mathbf{y} \). In this context, the observation space $\mathbb{R}^m$ serves as the \textit{ambient space} for the market data. Since the G2++ model is defined by five parameters, the set of all possible price vectors it can generate forms a 5-dimensional curved model surface embedded within $\mathbb{R}^m$. Calibration is geometrically equivalent to finding the point \( \hat{\mathbf{y}}_{\text{true}} = \mathbf{g}(\hat{\Pi}) \) on this manifold that minimizes the weighted Euclidean distance to the market data point \( \mathbf{y} \).

To analyze the stability of this solution, we linearize the model at the solution point. At \( \hat{\mathbf{y}}_{\text{true}} \), we construct a 5-dimensional \textit{tangent plane} spanned by the columns of the Jacobian matrix \( \mathbf{J}(\hat{\Pi}) \). 

The symmetric Weighted Hat Matrix $\mathbf{H}_w = \mathbf{W}^{1/2}\mathbf{J}(\mathbf{J}^\top \mathbf{W} \mathbf{J})^{-1}\mathbf{J}^\top\mathbf{W}^{1/2}$ defined in Section~\ref{subsec:geometry} acts as the true orthogonal projector in the variance-stabilized (scaled) space. When mapped back to the original ambient space, this operation acts as an \textit{affine projector} that maps the raw market data \( \mathbf{y} \) onto the tangent plane. The resulting point, \( \hat{\mathbf{y}}_{\text{fitted}} \), is the closest point on the \textit{linearized model} to the market data. The corresponding affine projection is given by
\begin{equation}
	\hat{\mathbf{y}}_{\text{fitted}} = \hat{\mathbf{y}}_{\text{true}} + \mathbf{W}^{-1/2} \mathbf{H}_w \mathbf{W}^{1/2} (\mathbf{y} - \hat{\mathbf{y}}_{\text{true}}) = \hat{\mathbf{y}}_{\text{true}} + \mathbf{J}(\mathbf{J}^\top \mathbf{W} \mathbf{J})^{-1} \mathbf{J}^\top \mathbf{W}(\mathbf{y} - \hat{\mathbf{y}}_{\text{true}}).
\end{equation}
The difference between the manifold point \( \hat{\mathbf{y}}_{\text{true}} \) and the tangent point \( \hat{\mathbf{y}}_{\text{fitted}} \) is a direct consequence of the model's nonlinearity. For a well-fitting model, this difference is small, which validates the use of linear diagnostic tools.

The validity of this linearized framework is tied to the asymptotic properties of the estimator. Specifically, the Noether condition (see \cite[Def. 2.3.6 and Thm. 2.4.4]{rieder1994robust}) requires the asymptotic negligibility of individual observations to ensure the normality of the estimator. In the context of our diagnostics, this is monitored via the leverage scores $h_{kk}$. As established in the literature on regression diagnostics \cite{cook1989regression, stahel1991directions}, a leverage score approaching $1$ indicates that the model is structurally forced to fit a specific data point, potentially violating the Noether condition and signaling a region of parameter instability.

\subsection{Pseudo-Inverse and Tikhonov EDoF}
\label{app:edof_regularization_comparison}

The interpretation of the Effective Degrees of Freedom in Section~\ref{subsec:dof} relies on the fact that the Weighted Hat Matrix is constructed using the Moore--Penrose pseudo-inverse. In that case,
\[
\mathbf{H}_w=\mathbf{W}^{1/2}\mathbf{J}\bigl(\mathbf{J}^\top\mathbf{W}\mathbf{J}\bigr)^{+}\mathbf{J}^\top\mathbf{W}^{1/2}
\]
is an orthogonal projector onto the column space of \(\mathbf{W}^{1/2}\mathbf{J}\), and therefore
\[
\operatorname{tr}(\mathbf{H}_w)=\operatorname{rank}(\mathbf{J}).
\]
As a consequence, the EDoF take exact integer values and can be interpreted as the number of locally active directions in parameter space.

A conceptually different alternative is Tikhonov regularization. Replacing the pseudo-inverse by
\[
\bigl(\mathbf{J}^\top\mathbf{W}\mathbf{J}+\lambda \mathbf{I}\bigr)^{-1}
\]
yields a regularized smoother matrix rather than a true projector. Its trace remains well defined, but it no longer coincides with the rank of the Jacobian and may take non-integer values.

\begin{figure}[htbp]
	\centering
	\begin{subfigure}[b]{\textwidth}
		\includegraphics[width=\textwidth]{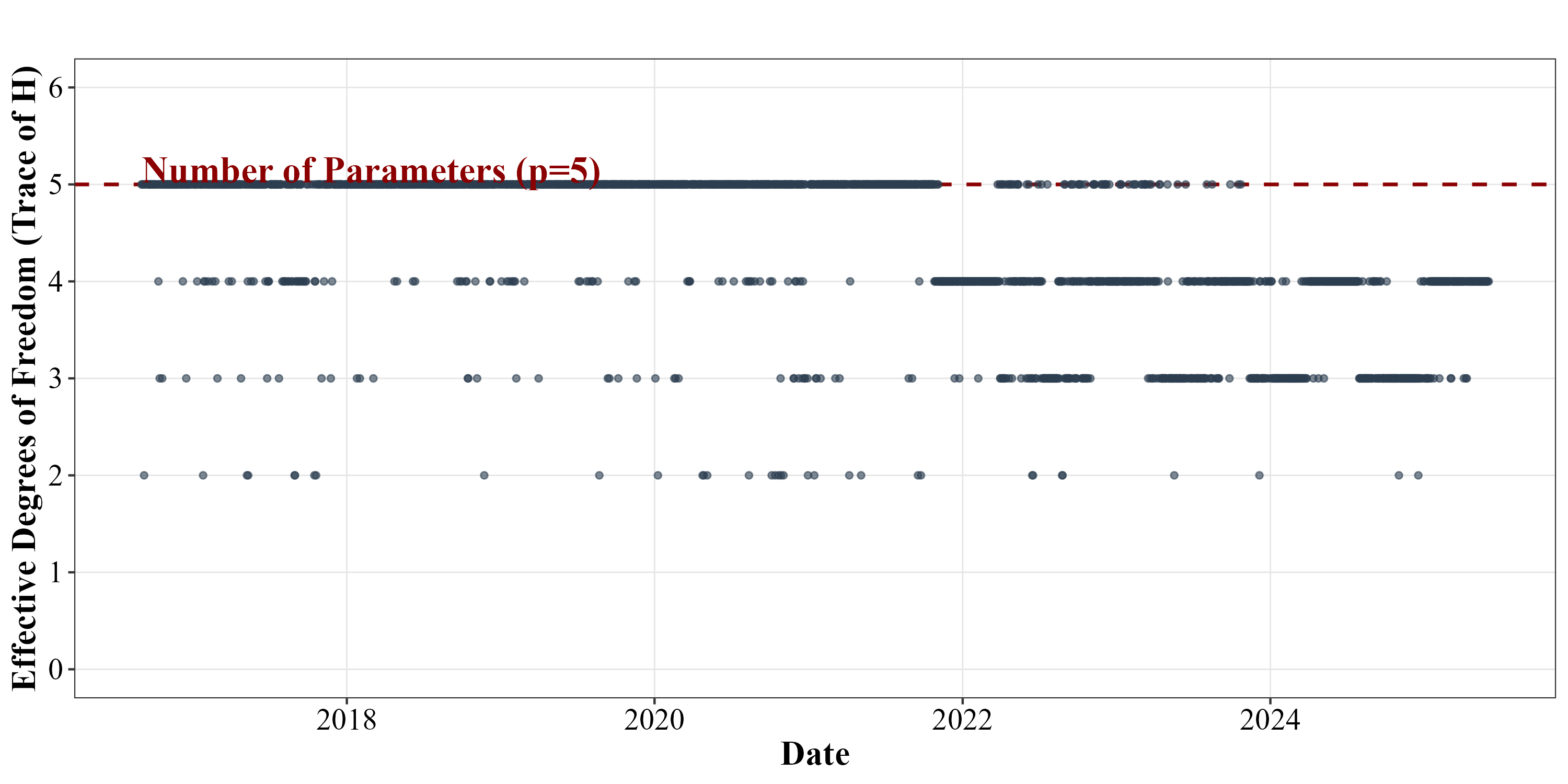}
		\caption{SVD-based pseudo-inverse}
	\end{subfigure}
	\hfill
	\begin{subfigure}[b]{\textwidth}
		\includegraphics[width=\textwidth]{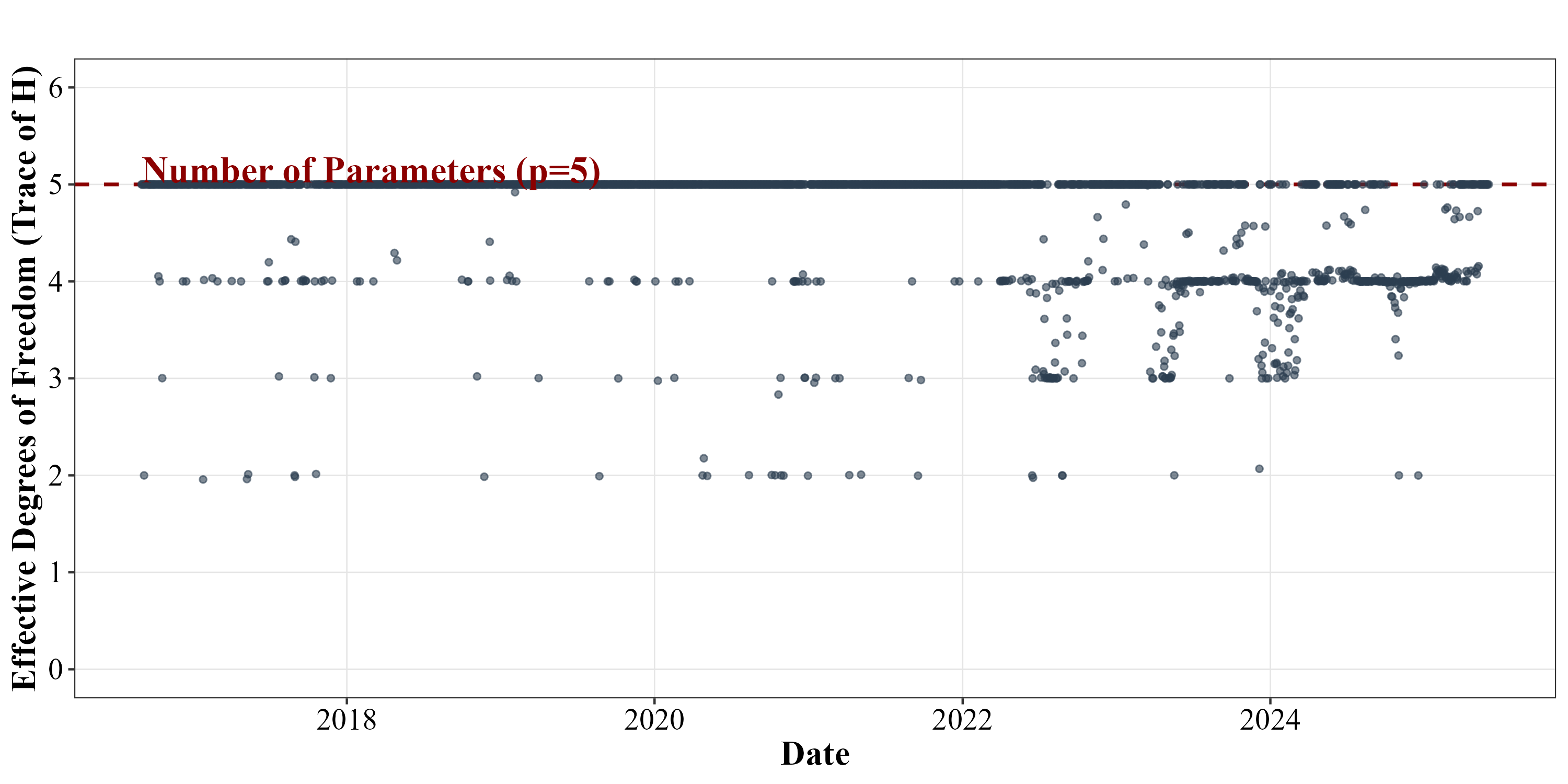}
		\caption{Tikhonov-regularized variant}
	\end{subfigure}
	\caption{\textbf{Effective Degrees of Freedom under two inversion schemes.} Panel (a) uses the SVD-based pseudo-inverse and therefore produces exact integer EDoF levels equal to the local rank of the Jacobian. Panel (b) uses a Tikhonov-type regularization, which replaces the projector by a regularized smoother and yields non-integer trace values.}
	\label{fig:edof_compare_regularization}
\end{figure}

Figure~\ref{fig:edof_compare_regularization} compares these two approaches. Under the pseudo-inverse construction, the EDoF fall on sharply separated integer bands, which makes rank loss directly visible and easy to interpret. Under Tikhonov regularization, by contrast, the EDoF appear as a range of non-integer values. This may be numerically convenient, but it makes the distinction between true local rank loss and regularization effects less transparent.

For the purposes of this paper, the pseudo-inverse-based formulation is preferable because the diagnostic is intended to identify genuine losses of local dimensionality. In that setting, integer-valued EDoF are central to an interpretable rank-based diagnostic.

\subsection{Variance-Stabilizing Transformations and the Delta Method}
\label{app:vst_theory}

While standard asymptotic theory implies normality for the WLS estimator, the G2++ parameters have strict physical boundaries (e.g., $|\rho_{xy}| \leq 1$ and $a_x, a_y, \sigma_x, \sigma_y > 0$). A standard linear confidence interval based on the raw covariance matrix can easily violate these boundaries in finite samples. To address this, we employ the univariate \textit{Delta Method} applied componentwise, combined with \textit{Variance-Stabilizing Transformations} (VST).

\paragraph{The Delta Method}
Let $\eta: \Theta \to \mathbb{R}$ be a smooth, invertible transformation applied to a scalar parameter estimate $\hat{\theta}_n$. By the standard Delta Method (see, e.g., \cite[Ch. 3]{van1998asymptotic}), the asymptotic distribution of the transformed estimator is:
\begin{equation}
	\sqrt{n}(\eta(\hat{\theta}_n) - \eta(\theta)) \xrightarrow{d} \mathcal{N}\left(0,[\eta'(\theta)]^2 V(\theta)\right),
\end{equation}
where $V(\theta)$ denotes the asymptotic variance of the original estimator $\hat{\theta}_n$.

\paragraph{Derivation of the Transformations}
We seek a specific transformation $\eta(\cdot)$ that stabilizes the variance, meaning it renders the asymptotic variance of the transformed parameter constant and independent of the true parameter value $\theta$. This is achieved by solving the differential equation $\eta'(\theta) \propto 1/\sqrt{V(\theta)}$, which yields:
\begin{equation}
	\eta(\theta) = \int \frac{1}{\sqrt{V(\theta)}} d\theta.
\end{equation}

\begin{itemize}
	\item \textbf{Case 1: Correlation ($\rho_{xy}$):} For correlation-type estimators, the asymptotic variance typically scales proportionally to $(1 - \rho^2)^2$. Substituting $V(\rho) \propto (1 - \rho^2)^2$ into the integral yields the \textit{Fisher z-transformation}:
	\begin{equation}
		\eta(\rho) = \int \frac{1}{1 - \rho^2} d\rho = \operatorname{arctanh}(\rho).
	\end{equation}
	
	\item \textbf{Case 2: Volatility and Mean Reversion ($\sigma, a$):} For strictly positive scale parameters, the variance typically scales with the square of the parameter magnitude ($V(\theta) \propto \theta^2$). The corresponding stabilizing solution is the \textit{logarithmic transformation}:
	\begin{equation}
		\eta(\theta) = \int \frac{1}{\theta} d\theta = \log(\theta).
	\end{equation}
\end{itemize}
These transformations map the bounded parameter domains to the entire real line $\mathbb{R}$. This ensures that symmetric confidence intervals constructed in the homoscedastic, transformed space strictly respect the natural boundaries of the parameters when mapped back to the original domain via the inverse transformation $\eta^{-1}$.

\subsection{Classical versus Robust Uncertainty Scaling}
\label{app:classical_vs_robust_ci}

To assess the sensitivity of the uncertainty quantification to the choice of residual scale estimator, we compare the classical MSE-based covariance scaling with the robust MAD-based alternative introduced in Section~\ref{subsec:uncertainty}. 

\begin{figure}[htbp]
	\centering
	\begin{subfigure}[b]{\textwidth}
		\includegraphics[width=\textwidth]{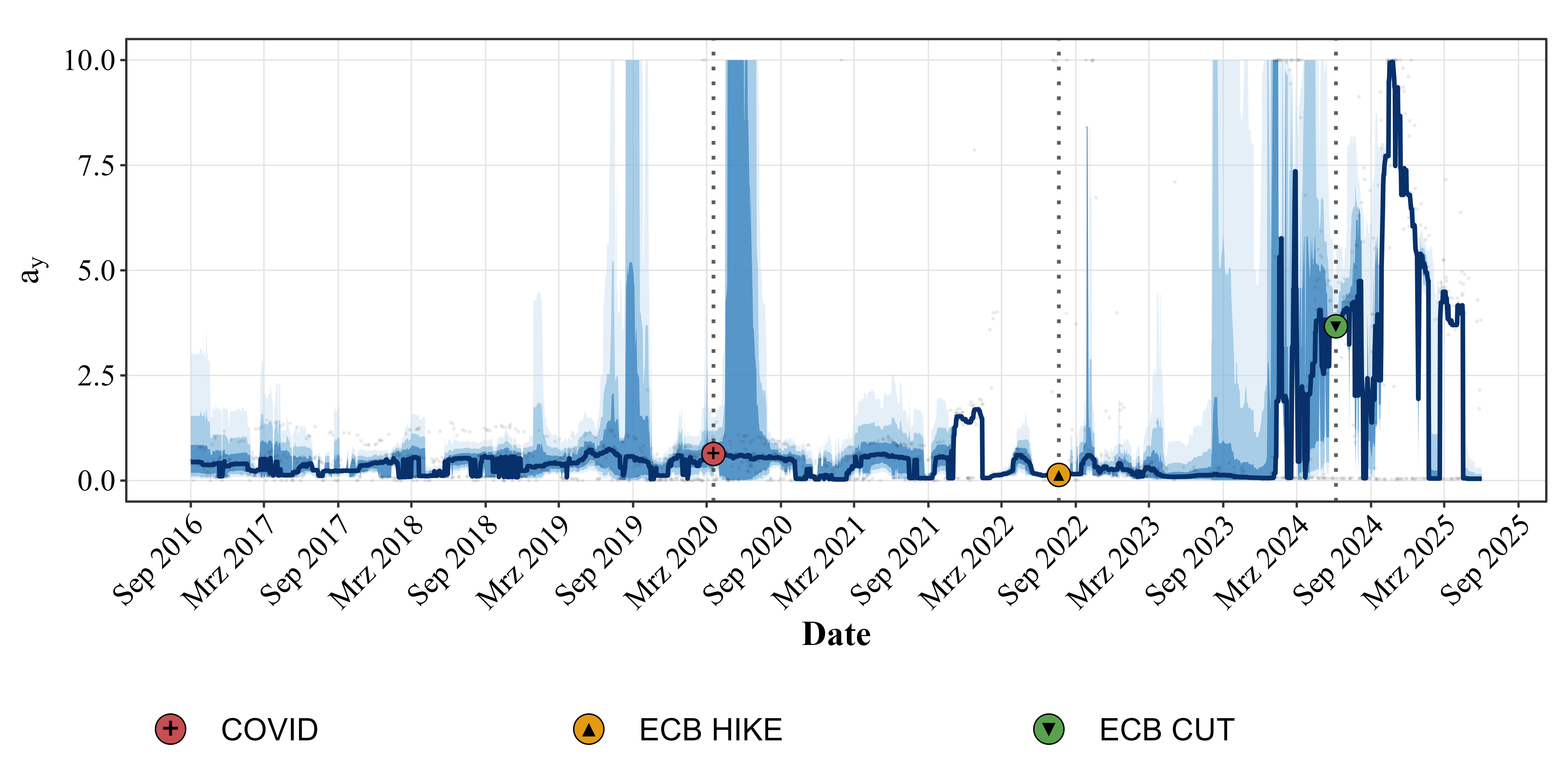}
		\caption{Classical MSE-based scaling}
	\end{subfigure}
	\hfill
	\begin{subfigure}[b]{\textwidth}
		\includegraphics[width=\textwidth]{fig07_robust_ci_ay.png}
		\caption{Robust MAD-based scaling}
	\end{subfigure}
	\caption{\textbf{Classical versus robust uncertainty scaling for $a_y$.} Panel (a) uses the classical MSE-based residual scale, while Panel (b) uses the robust MAD-based scale.}
	\label{fig:ci_compare_ay}
\end{figure}

Figure~\ref{fig:ci_compare_ay} illustrates the practical difference for the parameter \(a_y\). The two fan charts are based on the same covariance framework, the same pseudo-inverse treatment of rank deficiency, and the same Delta Method propagation. They differ only in the scaling of the covariance matrix. The classical MSE-based version reacts much more strongly to a small number of extreme residuals, especially at the short end of the cap surface. As a result, its uncertainty bands are wider than those of the robust alternative and in several episodes extend across the full admissible parameter range.

The same three reference dates as in the main text are included in both panels. They serve only as common reference points across the diagnostic plots, not as a separate event study. The comparison shows that the choice of residual scale affects the width and persistence of the uncertainty bands independently of a separate event interpretation: the classical MSE-based scaling produces broader and sometimes full-range bands, whereas the robust MAD-based scaling localizes broad uncertainty episodes more strongly.

This comparison makes the \textit{masking effect} directly visible in the sense of classical regression diagnostics; see Atkinson~\cite{Atkinson1986Masking}. Under classical scaling, a small number of large residuals can dominate the global MSE-based scale estimate and thereby inflate the confidence bands over extended periods, masking the typical uncertainty level on most days. The robust MAD-based scaling mitigates this effect by basing the covariance scaling on a robust measure of the typical residual size rather than on the most extreme deviations; compare also Croux, Dhaene, and Hoorelbeke~\cite{CrouxDhaeneHoorelbeke2004} on robust standard errors for robust estimators.

At the same time, both panels also display episodes in which the confidence interval for \(a_y\) collapses to the point estimate. These episodes are not caused by the choice of residual scale, but by the pseudo-inverse treatment of singular $\mathbf{J}^\top \mathbf{W}\mathbf{J}$. When rank is lost, the covariance approximation is computed only on the identified subspace, and discarded directions receive zero variance in the pseudo-inverse. Hence, such episodes should not be interpreted as unusually high statistical precision. Rather, they signal a degeneracy of the local covariance approximation and must be read jointly with the Effective Degrees of Freedom diagnostic.

\end{appendices}

\FloatBarrier


\clearpage
\begin{thebibliography}{23}
\providecommand{\natexlab}[1]{#1}
\providecommand{\url}[1]{\texttt{#1}}
\expandafter\ifx\csname urlstyle\endcsname\relax
  \providecommand{\doi}[1]{doi: #1}\else
  \providecommand{\doi}{doi: \begingroup \urlstyle{rm}\Url}\fi

\bibitem[Gneiting(2011)]{gneiting2011making}
Tilmann Gneiting.
\newblock Making and evaluating point forecasts.
\newblock \emph{Journal of the American Statistical Association}, 106\penalty0
  (494):\penalty0 746--762, 2011.
\newblock \doi{10.1198/jasa.2011.r10138}.

\bibitem[Duffie and Kan(1996)]{duffie1996yield}
Darrell Duffie and Rui Kan.
\newblock A yield-factor model of interest rates.
\newblock \emph{Mathematical Finance}, 6\penalty0 (4):\penalty0 379--406, 1996.
\newblock \doi{10.1111/j.1467-9965.1996.tb00123.x}.

\bibitem[Filipovi{\'c}(2009)]{filipovic2009term}
Damir Filipovi{\'c}.
\newblock \emph{Term-Structure Models: A Graduate Course}.
\newblock Springer Finance. Springer Berlin, Heidelberg, 2009.
\newblock ISBN 9783540680154.
\newblock \doi{10.1007/978-3-540-68015-4}.

\bibitem[Brigo and Mercurio(2001)]{brigoMercurio2001}
Damiano Brigo and Fabio Mercurio.
\newblock \emph{Interest Rate Models: Theory and Practice}.
\newblock Springer Finance. Springer, Berlin, Heidelberg, 2001.
\newblock ISBN 978-3-540-41772-9.

\bibitem[Wolf et~al.(2026)Wolf, Meyer, Mahler, and Diehl]{wolf2026openirm}
Mark-Oliver Wolf, Benedict~Nikolaus Meyer, Philipp Mahler, and Maximilian
  Diehl.
\newblock {openIRM}: Publicly accessible internal risk model of an artificial
  life insurer for analyzing and benchmarking actuarial methods in the
  {Solvency II} setting.
\newblock \emph{European Actuarial Journal}, 16:\penalty0 225--282, 2026.
\newblock \doi{10.1007/s13385-025-00435-6}.

\bibitem[Cont and Tankov(2003)]{cont2004financial}
Rama Cont and Peter Tankov.
\newblock \emph{Financial Modelling with Jump Processes}.
\newblock Chapman \& Hall/CRC Financial Mathematics Series. Chapman \&
  Hall/CRC, Boca Raton, FL, 2003.
\newblock ISBN 9781584884132.
\newblock \doi{10.1201/9780203485217}.

\bibitem[Rainer(2009)]{rainer2009calibration}
Martin Rainer.
\newblock Calibration of stochastic models for interest rate derivatives.
\newblock \emph{Optimization}, 58\penalty0 (3):\penalty0 373--388, 2009.
\newblock \doi{10.1080/02331930902741796}.

\bibitem[Karlsson et~al.(2017)Karlsson, Pilz, and
  Schl{\"o}gl]{karlsson2017calibrating}
Patrik Karlsson, K.~F. Pilz, and Erik Schl{\"o}gl.
\newblock Calibrating a market model with stochastic volatility to commodity
  and interest rate risk.
\newblock \emph{Quantitative Finance}, 17\penalty0 (6):\penalty0 907--925,
  2017.
\newblock \doi{10.1080/14697688.2016.1254814}.

\bibitem[Christoffersen and Jacobs(2004)]{christoffersen2004importance}
Peter Christoffersen and Kris Jacobs.
\newblock The importance of the loss function in option valuation.
\newblock \emph{Journal of Financial Economics}, 72\penalty0 (2):\penalty0
  291--318, 2004.
\newblock \doi{10.1016/j.jfineco.2003.02.001}.

\bibitem[Desmettre and Korn(2018)]{desmettre2018moderne}
Sascha Desmettre and Ralf Korn.
\newblock \emph{Moderne Finanzmathematik -- Theorie und praktische Anwendung
  Band 2: Erweiterungen des {Black--Scholes}-Modells, Zins, Kreditrisiko und
  Statistik}.
\newblock Studienb{\"u}cher Wirtschaftsmathematik. Springer Spektrum,
  Wiesbaden, 2018.
\newblock ISBN 9783658209995.
\newblock \doi{10.1007/978-3-658-21000-7}.

\bibitem[Seber and Wild(1989)]{seber1989nonlinear}
George A.~F. Seber and Christopher~J. Wild.
\newblock \emph{Nonlinear Regression}.
\newblock Wiley Series in Probability and Statistics. Wiley, New York, 1989.
\newblock ISBN 9780471617600.
\newblock \doi{10.1002/0471725315}.

\bibitem[Huber(1981)]{huber1981robust}
Peter~J. Huber.
\newblock \emph{Robust Statistics}.
\newblock John Wiley \& Sons, New York, 1981.

\bibitem[Hampel et~al.(1986)Hampel, Ronchetti, Rousseeuw, and
  Stahel]{hampel1986robust}
Frank~R. Hampel, Elvezio~M. Ronchetti, Peter~J. Rousseeuw, and Werner~A.
  Stahel.
\newblock \emph{Robust Statistics: The Approach Based on Influence Functions}.
\newblock John Wiley \& Sons, New York, 1986.

\bibitem[Reeds(1976)]{reeds1976definition}
James~A. Reeds.
\newblock \emph{On the Definition of von Mises Functionals}.
\newblock PhD thesis, Harvard University, Cambridge, MA, 1976.

\bibitem[Rieder(1994)]{rieder1994robust}
Helmut Rieder.
\newblock \emph{Robust Asymptotic Statistics}.
\newblock Springer Series in Statistics. Springer, New York, 1994.

\bibitem[Ben-Israel and Greville(2003)]{BenIsraelGreville2003}
Adi Ben-Israel and Thomas N.~E. Greville.
\newblock \emph{Generalized Inverses: Theory and Applications}.
\newblock CMS Books in Mathematics. Springer, New York, 2 edition, 2003.
\newblock ISBN 9780387002934.
\newblock \doi{10.1007/b97366}.

\bibitem[Hintze and Nelson(1998)]{HintzeNelson1998}
Jerry~L. Hintze and Ray~D. Nelson.
\newblock Violin plots: A box plot-density trace synergism.
\newblock \emph{The American Statistician}, 52\penalty0 (2):\penalty0 181--184,
  1998.
\newblock \doi{10.1080/00031305.1998.10480559}.

\bibitem[{van der Vaart}(1998)]{van1998asymptotic}
Aad~W. {van der Vaart}.
\newblock \emph{Asymptotic Statistics}.
\newblock Cambridge Series in Statistical and Probabilistic Mathematics.
  Cambridge University Press, Cambridge, 1998.
\newblock ISBN 9780521496032.
\newblock \doi{10.1017/CBO9780511802256}.

\bibitem[Molnar(2022)]{molnar2022interpretable}
Christoph Molnar.
\newblock \emph{Interpretable Machine Learning: A Guide for Making Black Box
  Models Explainable}.
\newblock Self-published, Munich, 2 edition, 2022.
\newblock URL \url{https://christophm.github.io/interpretable-ml-book/}.

\bibitem[Cook and Weisberg(1989)]{cook1989regression}
R.~Dennis Cook and Sanford Weisberg.
\newblock Regression diagnostics with dynamic graphics.
\newblock \emph{Technometrics}, 31\penalty0 (3):\penalty0 277--291, 1989.
\newblock \doi{10.1080/00401706.1989.10488547}.

\bibitem[Stahel and Weisberg(1991)]{stahel1991directions}
Werner~A. Stahel and Sanford Weisberg, editors.
\newblock \emph{Directions in Robust Statistics and Diagnostics: Part II},
  volume~34 of \emph{The IMA Volumes in Mathematics and its Applications}.
\newblock Springer, New York, 1991.
\newblock ISBN 978-0-387-97531-3.
\newblock \doi{10.1007/978-1-4612-4444-8}.

\bibitem[Atkinson(1986)]{Atkinson1986Masking}
A.~C. Atkinson.
\newblock Masking unmasked.
\newblock \emph{Biometrika}, 73\penalty0 (3):\penalty0 533--541, 1986.
\newblock \doi{10.1093/biomet/73.3.533}.

\bibitem[Croux et~al.(2003)Croux, Dhaene, and
  Hoorelbeke]{CrouxDhaeneHoorelbeke2004}
Christophe Croux, Geert Dhaene, and Dirk Hoorelbeke.
\newblock Robust standard errors for robust estimators.
\newblock CES Discussion Paper DPS 03.16, CES, KU Leuven, 2003.

\end{thebibliography}
\end{document}